\documentclass[longauth]{aa}  

\usepackage{graphicx}
\usepackage{txfonts}
\usepackage{xcolor}
\usepackage{lipsum}
\usepackage{subcaption}         % necessary for continued figures, example in section 3
\usepackage{lscape}             % to rotate a single page table, example in appendix.
\usepackage{placeins}           % useful with \FloatBarrier, to keep 
\usepackage{hyperref}    
\graphicspath{ {Figures/} }
                                
\begin{document}

   \title{TOI-159~b: an eccentric hot-Jupiter planet around a young, \\pulsating $\gamma$~Doradus star}

   \subtitle{}

   \author{G. Mantovan
          \inst{\ref{inst1},\ref{inst2}}$^{\orcid{0000-0002-6871-6131}}$
          \and
          A. Llancaqueo Albornoz\inst{\ref{inst3}}
          \and
          A. Psaridi\inst{\ref{inst4}, \ref{inst5}, \ref{inst6}}$^{\orcid{0000-0002-4797-2419}}$
          \and
          A. Thompson\inst{\ref{inst7}}$^{\orcid{0000-0003-4128-2270}}$
          \and
          T. Zingales\inst{\ref{inst8},\ref{inst2}}$^{\orcid{0000-0001-6880-5356}}$
          \and
          V. Nascimbeni\inst{\ref{inst2}}$^{\orcid{0000-0001-9770-1214}}$
          \and
          S. Villanova\inst{\ref{inst9}}
          \and
          G. Piotto\inst{\ref{inst8},\ref{inst2}}$^{\orcid{0000-0002-9937-6387}}$
          \and
          K. A.\ Collins\inst{\ref{inst10}}$^{\orcid{0000-0001-6588-9574}}$
          \and
          J. Serna\inst{\ref{inst11}}$^{\orcid{0000-0001-7351-6540}}$
          \and
          L. Malavolta\inst{\ref{inst6},\ref{inst2}}$^{\orcid{0000-0002-6492-2085}}$
          \and
          K. Stassun\inst{\ref{inst12}}$^{\orcid{0000-0002-3481-9052}}$
          \and
          F. Bouchy\inst{\ref{inst6}}$^{\orcid{0000-0002-7613-393X}}$
          \and
          C. C. Cortés\inst{\ref{inst13}}$^{\orcid{0000-0002-5003-3762}}$
          \and
          P. Evans\inst{\ref{inst15}}
          \and
          T. Gan\inst{\ref{inst16}}$^{\orcid{0000-0002-4503-9705}}$
          \and
          M. Lendl\inst{\ref{inst6}}$^{\orcid{0000-0001-9699-1459}}$
          \and
          M. B. Lund\inst{\ref{inst17}}$^{\orcid{0000-0003-2527-1598}}$
          \and
          D. Nardiello\inst{\ref{inst8},\ref{inst2}}$^{\orcid{0000-0003-1149-3659}}$
        }

   \institute{Centro di Ateneo di Studi e Attivit\`a Spaziali ``G. Colombo'' -- Universit\`a di Padova, Via Venezia 15, IT-35131, Padova, Italy\\ \email{giacomo.mantovan@unipd.it}\label{inst1}
            \and
             Istituto Nazionale di Astrofisica - Osservatorio Astronomico di Padova, Vicolo dell'Osservatorio 5, IT-35122, Padova, Italy\label{inst2}
             \and
              Universidad de Concepción\label{inst3}
              \and
              Institute of Space Sciences (ICE, CSIC), Carrer de Can Magrans S/N, Campus UAB, Cerdanyola del Valles, E-08193, Spain\label{inst4}
                \and
                Institut d’Estudis Espacials de Catalunya (IEEC), 08860 Castelldefels (Barcelona), Spain\label{inst5}
                \and
              Observatoire Astronomique de l'Université de Genève, Chemin Pegasi 51, CH-1290 Versoix, Switzerland\label{inst6}
              \and
              Department of Physics and Astronomy, University College London, Gower Street, WC1E 6BT London, UK\label{inst7}
                \and
             Dipartimento di Fisica e Astronomia ``Galileo Galilei'', Università di Padova, Vicolo dell'Osservatorio 3, IT-35122, Padova, Italy
              \label{inst8}
              \and
              Universidad Andres Bello, Facultad de Ciencias Exactas, Departamento de F{\'i}sica y Astronom{\'i}a - Instituto de Astrof{\'i}sica, Autopista Concepci\'on-Talcahuano 7100, Talcahuano, Chile\label{inst9}
              \and
              Center for Astrophysics \textbar \ Harvard \& Smithsonian, 60 Garden Street, Cambridge, MA 02138, USA\label{inst10}
              \and
              Homer L. Dodge Department of Physics and Astronomy, University of Oklahoma, Norman, OK 73019, USA\label{inst11}
              \and 
              Department of Physics and Astronomy, Vanderbilt University, Nashville, TN 37235, USA\label{inst12} 
              \and
              Centro de Investigación en Ciencias del Espacio y Física Teórica, Universidad Central de Chile, Av. Francisco de Aguirre 0405, La Serena, Chile\label{inst13}
              \and
              Phil Evans, El Sauce Observatory, Coquimbo Province, Chile\label{inst15}
              \and 
              Department of Astronomy and Tsinghua Centre for Astrophysics, Tsinghua University, Beijing 100084, China\label{inst16}
              \and
              NASA Exoplanet Science Institute, IPAC, California Institute of Technology, Pasadena, CA 91125, USA\label{inst17}
             }

    \date{Compiled: \today}
 
  \abstract
   {Fast-rotating hot stars are challenging targets for exoplanet searches due to rotational broadening and stellar variability. Moreover, hot stars often exhibit pulsations, an additional source of scatter in both photometric and spectroscopic series. Because of these challenges, such stars remain a relatively unexplored environment for planetary architecture and evolution studies. In this study, we present the confirmation and preliminary atmospheric characterisation of a giant planet orbiting a young ($\approx$ 150 Myr), pulsating $\gamma$ Doradus star. TOI-159~b ($P_{\rm orb} \simeq 3.7$ d, $R_{\rm p} \simeq 1.6~R_{\rm J}$, $M_{\rm p} \simeq 3.5 M_{\rm J}$) is an S-type planet in a close binary system and is the hottest ($T_{\rm eq} \simeq 1900$ K) hot Jupiter with a significant eccentricity ($e = 0.24 \pm 0.04$) ever detected. Our joint modelling of radial velocities (HARPS and CORALIE), transits (\textit{TESS}), and spectro-photometry (IMACS) allows us to detect its Keplerian signal at high significance ($13 \sigma$), place strong constraints on its eccentricity ($6 \sigma$), disentangle the stellar rotational modulation and pulsation periods, and generate a low-resolution transmission spectrum, on which we conduct an exploratory analysis to constrain the presence of a planetary atmosphere using combined star-planet retrievals. Whilst our spectrum appears to display some modulation, the data is too coarse to allow for any conclusive detections at this stage. Higher-resolution observations are needed to confirm or refute these features and, if genuine, determine whether they originate from contamination from the star or a planetary atmosphere. }


   \maketitle

%%%%%%%%%%%%%%%%%%%%%%%%%%%%%%%%%%%%%%%%%%%%%%%%%%%%%%%%%%%%%%
\section{Introduction}
Pulsating stars are remarkable targets because their oscillations offer a unique opportunity to probe their internal properties and processes \citep{2010aste.book.....A}. In fact, the stellar properties affect the frequencies of these oscillations, which we can determine via time series analysis. For instance, the calculation of the amplitude spectrum \citep{1976Ap&SS..39..447L, 1982ApJ...263..835S} allows us to find the frequencies associated with the pulsations. The study of stellar oscillations can thus provide important constraints on stellar structures and evolutionary models \citep{2021RvMP...93a5001A}, allowing us to better characterise stellar parameters such as mass, age, distance, and rotation \citep{2019ARA&A..57...35A}. Of particular interest are stars that exhibit nonradial oscillations, as they can reveal details about the deepest layers of the stellar interior. Among such stars are the $\gamma$ Doradus ($\gamma$ Dor) pulsators, whose variations are driven by high-order, nonradial, gravity-mode pulsations \citep{2005A&A...435..927D}. Due to the combination of their low-amplitude photometric variability and radial velocity variations up to a few km s$^{-1}$ \citep[e.g.][]{2011AJ....142...39H}, the discovery of exoplanets around them is usually hindered by detection biases and limited to imaging detections \citep{2008Sci...322.1348M,2024AJ....167..205T}. Nevertheless, the number of confirmed transiting exoplanets orbiting $\gamma$ Dor stars is rapidly increasing \citep[e.g.][]{2010MNRAS.407..507C,2015AJ....150...85H,2023A&A...674A..22G}.

Transiting exoplanets also offer the opportunity to probe their atmospheres. In fact, if the planet has an atmosphere, the opacity will gradually decrease with height as the atmosphere becomes more tenuous \citep{2006A&A...448..379E}. The stellar light passing through the atmosphere is selectively absorbed by atoms and molecules as a function of $\lambda$, or scattered by Mie or Rayleigh processes. As a consequence, the observed wavelength dependence of the transit depth can be used to probe the planet atmospheric composition. The ``effective'' optical planet radius is in general a function of wavelength $R_{\rm p}(\lambda)$, known as a \textit{transmission spectrum}. From such spectra we can detect well-defined signatures from atoms/molecules \citep[see][for a review]{2010ARA&A..48..631S} and get valuable information on the metal-enrichment of the planetary atmosphere \citep[e.g.][]{2011ApJ...736L..29M, 2016ApJ...831...64T}. Atomic metals have been observed in the UV and optical transmission spectra, with previous detections of neutral Na and K \citep{2002ApJ...568..377C, 2021JGRE..12606629F} due to these species having particularly prominent features at 589 and 770~nm, respectively. Moreover, we can detect scattering by haze particles \citep{2011MNRAS.416.1443S} by measuring the slope of the spectral continuum. We can constrain the atmospheric scale height $H$, and consequently the mean molecular weight ($\mu$) of the atmosphere by measuring the transmission spectrum and detecting atmospheric scattering and/or absorption in it, and by assuming an equilibrium temperature $T_{\rm eq}$, or vice versa \citep{2008A&A...481L..83L}. Eventually, through theoretical models \citep[e.g.][]{2013A&A...559A..32N} we can cast light on the atmospheric composition of the planet and also on its inner structure and evolutionary history. 

We need high-precision radial velocities (RVs) to derive masses and eccentricities, which, combined with the radii from transit, yield reliable inner bulk densities estimates and enable us to explore potential differences in planetary structure and evolution of giant exoplanets. This accurate mass determination is crucial to characterise the atmosphere of exoplanets, because an accurate determination of the planetary surface gravity can lead to a better description of the atmospheric properties. Characterising atmospheres is essential for enhancing our understanding of the origins of giant planets \citep{2021JGRE..12606629F}.

In this paper we confirm and characterise a hot Jupiter (HJ) orbiting a $\gamma$ Dor pulsator using a combination of \textit{TESS} photometry, ground-based photometry, and RVs collected with the High Accuracy Radial velocity Planet Searcher (HARPS, \citealt{2003Msngr.114...20M}) spectrograph (Sect. \ref{sec:obs}). In Sect. \ref{sec:sel}, we describe the target selection and statistical validation of the HJ characterised in the present work, which was first identified by the \textit{TESS} pipelines. Section \ref{sec:stellar} describes the stellar properties determined by two independent methods. In Sect. \ref{sec:analysis} we show how we identified and confirmed the giant planet by outlining the joint modelling of photometry and spectroscopy. We then present the atmospheric analysis of TOI-159~b using the Magellan/IMACS transit observation. Section \ref{sec:discussion} discusses our results and presents the characteristics of the inflated, eccentric hot Jupiter TOI-159~b. Concluding remarks are given in Sect. \ref{sec:conclusions}. 

%%%%%%%%%%%%%%%%%%%%%%%%%%%%%%%%%%%%%%%%%%%%%%%%%%%%%%%%%%%%%%
\section{Target Selection}
\label{sec:sel}

The Transiting Exoplanet Survey Satellite (\textit{TESS}, \citealt{2015JATIS...1a4003R}) is an all-sky survey that looks for transiting exoplanets orbiting bright stars by delivering a large number of high-precision light curves. These data are inspected with several transit-search pipelines: the Quick-Look Pipeline (QLP, \citealt{2020RNAAS...4..204H,2020RNAAS...4..206H}), the Faint Star QLP Search pipeline at MIT \citep{2021RNAAS...5..234K}, and the Science Processing Operations Center \citep[SPOC,][]{2016SPIE.9913E..3EJ} at NASA Ames Research Center. Targets that show convincing transit-like signals are classified as \textit{TESS} Objects of Interest (TOIs, \citealt{2021ApJS..254...39G}).

From the full list of TOIs, we identified giant planets around bright, main sequence stars (TESS magnitude  $T_\mathrm{mag}< 12$), which were suitable for atmospheric characterisation and with transits observable from Las Campanas Observatory in Chile. In this paper, we show the confirmation and atmospheric characterisation of the HJ TOI-159.01. We emphasise that this target was predicted to be amenable for precise atmospheric characterisation with JWST because its Transmission Spectroscopy Metric (TSM, predicted value in Sect. \ref{sec:tsm}) is greater than 90 \citep{2018PASP..130k4401K}. The ground-based transmission spectrum using Magellan/IMACS optical transit is complementary to JWST IR transits and, therefore, crucial to better plan and prepare these future observations. 

%%%%%%%%%%%%%%%%%%%%%%%%%%%%%%%%%%%%%%%%%%%%%%%%%%%%%%%%%%%%%%
\section{Observations and data reduction}
\label{sec:obs}

\subsection{TESS photometry}\label{sec:tess}
The planet described in this work was recognised as a transiting planet candidate in \textit{TESS} photometry. \textit{TESS} observed TOI-159 (TIC 394657039) at 30~min cadence in Sectors 1, 2, 12, and at 2~min cadence in Sectors 13, 28, 32, 39, 62, 66, 68, 89, 93, 94, and 95. Table \ref{table:tess-obs} summarises the \textit{TESS} observations.

\begin{table}
\caption{Observations from \textit{TESS} summarised.}             
\label{table:tess-obs}      
\centering                          
\begin{tabular}{c c c}        
\hline\hline                
Sectors & Source & Cadence \rule{0pt}{2.5ex} \rule[-1ex]{0pt}{0pt} \\     
\hline                        
  1, 2 & TESS-SPOC & 1800~s  \rule{0pt}{2.5ex} \rule[-1ex]{0pt}{0pt}\\ 
  12 & QLP & 1800~s  \rule{0pt}{2.5ex} \rule[-1ex]{0pt}{0pt}\\ 
  13, 28, 32, 39, 62, 66, 68 & SPOC & 120~s  \rule{0pt}{2.5ex} \rule[-1ex]{0pt}{0pt}\\
  89, 93, 94, 95 & SPOC & 120~s  \rule{0pt}{2.5ex} \rule[-1ex]{0pt}{0pt}\\
\hline                                  
\end{tabular}
\end{table}

Two-minute cadence light curves were reduced by the TESS SPOC pipeline. We used Presearch Data Conditioning Simple Aperture Photometry (PDCSAP; \citealt{2012PASP..124.1000S,2012PASP..124..985S,2014PASP..126..100S}) SPOC light curves, which are already  corrected for systematic effects. 

The transit signals passed each \textit{TESS} data validation test and the \textit{TESS} Science Office issued an alert for TOI-159.01 on September 5, 2018. The SPOC pipeline later detected the transit signal of TOI-159 in the TESS-SPOC light curves. In addition, the SPOC Data Validation module \citep{2018PASP..130f4502T} performed the difference image centroiding test and constrained the location of the host star to be within $0.9 \pm 2.5$ arcsec of the transit source. Figure \ref{fig:159lc} shows the \textit{TESS} photometric time series.

\subsection{Ground-based Lightcurve Photometry\label{sec:groundtransits}}
The \textit{TESS} pixel scale is $\sim 21\arcsec$ pixel$^{-1}$ and photometric apertures typically extend out to roughly 1 arcminute, generally causing multiple stars to blend in the photometric aperture. To determine the true source of the \textit{TESS} detections, we acquired ground-based time-series follow-up photometry of the field around TOI-159 as part of the \textit{TESS} Follow-up Observing Program \citep[TFOP;][]{collins:2019}\footnote{https://tess.mit.edu/followup}. We used the {\tt TESS Transit Finder}, which is a customized version of the {\tt Tapir} software package \citep{Jensen:2013}, to schedule our transit observations. The light curve data are available on the {\tt EXOFOP-TESS} website\footnote{\href{https://exofop.ipac.caltech.edu/tess/target.php?id=394657039}{https://exofop.ipac.caltech.edu/tess/target.php?id=394657039}}. The lightcurves confirm the event on-target and show no strong transit depth chromaticity, but are not included in the global modeling.

\subsubsection{LCOGT 0.4\,m} 

We observed a full transit window of TOI-159.01, with limited post-egress baseline, in Sloan $i'$ band simultaneously using two telescopes on UTC 2018 November 15 from the Las Cumbres Observatory Global Telescope \citep[LCOGT;][]{Brown:2013} 0.4\,m network node at Cerro Tololo Inter-American Observatory (CTIO) in Chile.  The 0.4\,m telescope was equipped with a $2048\times3072$ SBIG STX6303 camera having an image scale of 0$\farcs$57 pixel$^{-1}$, resulting in a $19\arcmin\times29\arcmin$ field of view. The images were calibrated by the standard LCOGT {\tt BANZAI} pipeline \citep{McCully:2018}, and photometric data were extracted using {\tt AstroImageJ} \citep{Collins:2017}. Circular apertures with radius $9\arcsec$ were used to extract the differential photometry. A transit with depth $\sim11$ ppt was confirmed on-target.

\subsubsection{El Sauce 0.36\,m}

We observed a full transit window of TOI-159.01 in B-band, with limited pre-ingress baseline, on UTC 2018 December 19 from the Evans 0.36\,m telescope at El Sauce Observatory in Coquimbo Province, Chile. The telescope was equipped with a $1536\times1024$ SBIG STT-1603-3 camera. The image scale was 1$\farcs$47 pixel$^{-1}$, resulting in an $18.8\arcmin\times12.5\arcmin$ field of view. The photometric data were extracted using {\tt AstroImageJ} with a circular $8\farcs$8 photometric aperture. A transit with depth $\sim11$ ppt was confirmed on-target, and shows no strong transit depth chromaticity compared to the LCOGT lightcurves in Sloan $i'$ band. The lack of strong chromaticity suggested a good candidate for radial velocity follow-up.

\subsection{Magellan/IMACS spectrophotometry}\label{sec:imacs}

The atmosphere of TOI-159.01 has never been characterised, and measuring its scale height $H$ would be fundamental to get a hint of the inner structure of this giant planet. Our primary goal is to constrain the presence of Rayleigh scattering -- possibly induced by a hazy or low-$\mu$ atmosphere -- by measuring the spectral slope of the continuum and to look for optical absorbers within the IMACS wavelength range such as Na and metal oxides and hydrides. This could further support or, be in contrast with, the metal-enrichment trend seen in giant exoplanets \citep{2016ApJ...831...64T}. For this reason we obtained a full transit observation of TOI-159.01 on evening date 2022-11-13 (Programme CN2022B-25, PI:~Villanova) with the Inamori-Magellan Areal Camera and Spectrograph \citep[IMACS,][]{2011PASP..123..288D} mounted at the Magellan-Baade 6.5-m telescope \citep{Shectman2003}. 

The large photon-collecting area and versatility of IMACS allowed us to measure the transit depth of TOI-159.01 at different wavelengths through differential spectrophotometry. Specifically, the multi-object spectrograph (MOS) capabilities of IMACS allowed us to simultaneously collect spectra of our target and several comparison stars, which is crucial to mitigate systematic errors of instrumental and telluric origin. We can then extract a transmission spectrum of the planetary atmosphere by splitting the wavelength range of IMACS into several spectral bins, integrating the signal collected within each bin, normalising with respect to the flux of the reference stars, and analysing the ``chromatic'' light curve from each bin separately. The final output is a low-resolution transmission spectrum, necessary if we want to detect the continuum. In fact, during the usual normalisation process of a high-resolution spectrum, the continuum of the transmission spectrum is lost \citep[e.g.][]{2018A&A...612A..53P}. 

We setup the camera in the  $f$/4 MOS configuration and designed a custom mask with several 5"$\times$12" long slits. One slit was positioned on the target, while the remaining ones were assigned to a set of reference stars having magnitude and colour similar to the target. The large field of view of the $f$/4 camera (15.4'$\times$15.4') enables us to observe several comparison stars and correct for common-mode systematic errors. Such strategy compensates for residual slit losses and allows us to optimally subtract the sky, thus performing accurate differential spectrophotometry. By choosing the 300 line mm$^{-1}$ grating at a blaze angle of 4.3$^{\circ}$ setup, we provide a wavelength coverage of 400–950 nm with the $f$/4 observations. This range includes the neutral NaI and KI lines. 

We gathered a time series of IMACS spectra covering a full transit of TOI-159.01, with a constant exposure time of 60~s, starting $\sim$1 hour before the first contact and ending 1 hour after the last contact. The two hours of off-transit photometry are mandatory to analyse and remove any systematic trend which can arise during the series \citep[e.g.][]{2013A&A...559A..32N}. The data reduction has been performed starting from raw spectra and using IRAF \citep{Iraf1986,1993ASPC...52..173T} and the \texttt{igapmos} package to perform an optimal extraction of the individual spectrum, after correcting for bias/flat/dark and calibrating the wavelengths through standard techniques. The spectral range was then split into 12 wavelength bins, avoiding any region covered by the CCD gaps in any among the target and the reference stars. The identified wavelength bins are shown in Fig.~\ref{fig:channels} in the Appendix. A set of eleven chromatic light curves was then built by summing the counts of the target spectrum in each bin, then dividing the value by the sum of the spectral counts of each reference star in the same bin, and repeating the process for each spectrum of the time series. The time stamps of each light curve points were assigned by extracting the JD value from the FITS header of the corresponding spectrum, shifting the time to the centre of the exposure, and then converting the value to the BJD-TDB time standard by following the prescription by \citet{Eastman2010}. At this stage we realised that a significant part of the pre-ingress phase has been impacted by very bad seeing, leaving uncorrectable systematic errors due to differential slit losses. This part was discarded from our analysis, but fortunately a small amount of pre-transit data points survived to allow us a proper normalisation of the light curve. We also realised that the target spectra including our two bluest channels (\#1 4000-4290~\AA{} and \#2 4375-4510~\AA{}) is contaminated by light beyond any effective correction, and are discarded as well. We are then left with nine channels, those numbered from \#3 to \#11 in Fig.~\ref{fig:channels}).

The light curves were then passed to the following step of the analysis, performed simultaneously with the other photometric and spectroscopic data and described in Section~\ref{sec:analysis}.

\subsection{CORALIE spectroscopic follow-up}
\label{sec:coralie}
TOI-159 was initially observed with the CORALIE spectrograph \citep{Queloz2001} on the Swiss 1.2-m Euler telescope at La Silla Observatory in Chile as part of the TFOP. In total, we collected nine spectra spanning from 2018~October~6 to November~7, each obtained with an exposure time of 2700~s. CORALIE is a high-resolution, fibre-fed echelle spectrograph with a resolving power of $R \sim 60{,}000$, fed by a $2\arcsec$ science fibre and equipped with a secondary fibre containing a Fabry-Pérot étalon for simultaneous wavelength calibration during the observations. The scientific spectra were processed with the standard CORALIE data reduction pipeline and cross-correlated with a binary mask representing the characteristic features of a star of the selected spectral type \citep{Baranne1996}. The strong stellar activity distorts the core of the average line profile, while stellar rotation broadens the line. Hence, we selected a half-window large enough to include the continuum when fitting the cross-correlation function (CCF) profile. Specifically, we used a A0 mask with an half-window of 150 km s$^{-1}$. This mask provides a higher CCF contrast compared to masks of later spectral types, improving the precision of the RV extraction.

\subsection{HARPS spectroscopic follow-up}
\label{sec:harps}
TOI-159 has been observed with HARPS \citep{2003Msngr.114...20M} at ESO's 3.6 m telescope between October 2021 and January 2022 (proposal 108.22LR, PI: A. Psaridi), collecting 20 spectra, with exposure times of 1800 s. The wavelength range of these spectra is 378-691 nm with a resolving power of R $\sim$ 115~000. The average uncertainty in the RV measurements was about 18 m s$^{-1}$, while the typical S/N at 5460$\AA$ was 53. Due to the nature of the star, we required higher-efficiency observations to ensure higher S/N and improved RV precision. Therefore, all observations were done using the High Efficiency Mode (EGGS), which provides a factor of $\sim$1.75 gain in flux compared to the standard High Accuracy Mode (HAM), using a 1.4 arcsec science fibre on the target and fibre B on the sky \citep[see][]{2003Msngr.114...20M}.

\begin{figure}[!t]
   \centering
   \includegraphics[width=\hsize]%
    {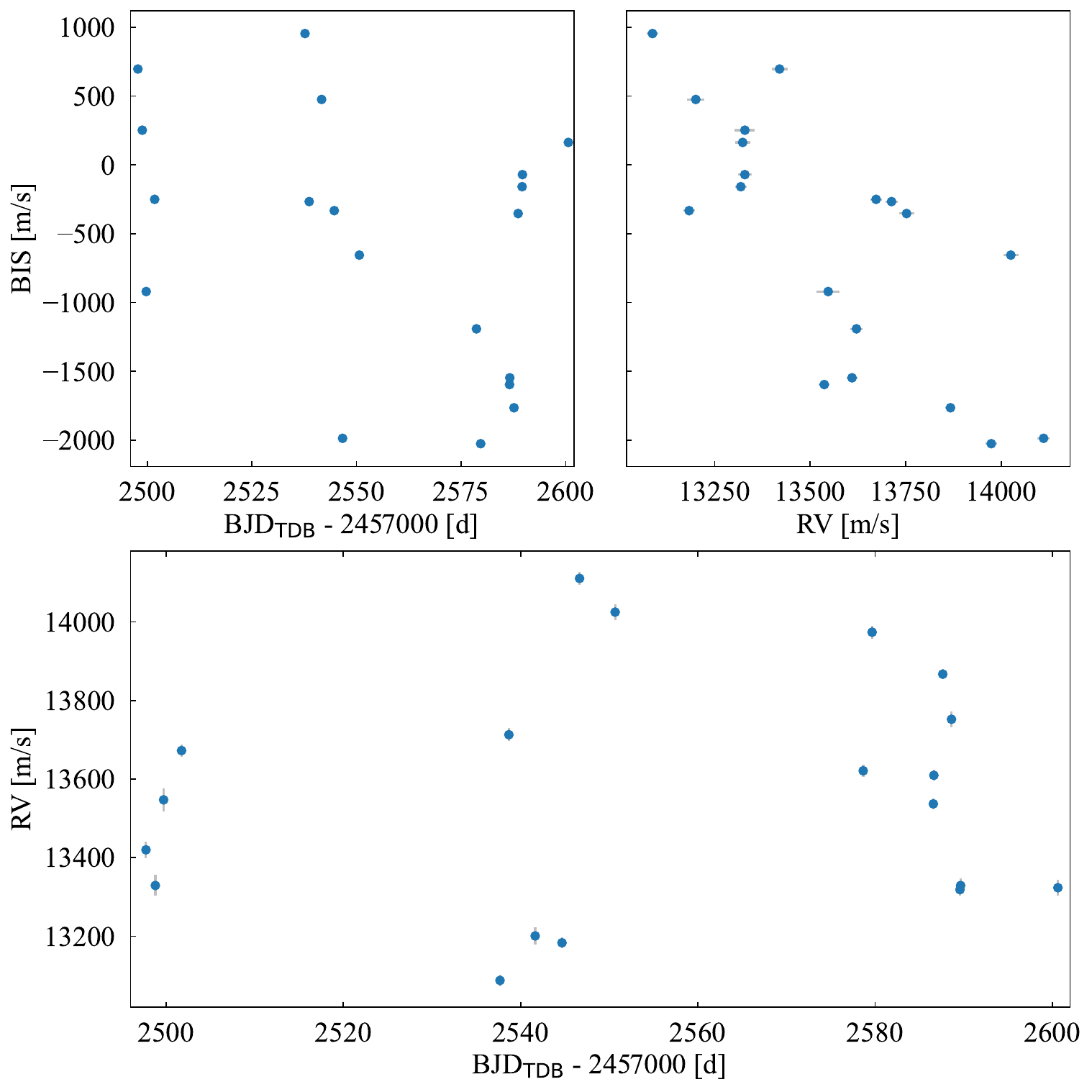}
   \caption{HARPS spectroscopic time series of TOI-159. \textit{Bottom:} RV time series. \textit{Top left:} BIS time series. \textit{Top right:} RV vs BIS time series, where we can see a clear correlation between them. }
   \label{fig:159_gls}
\end{figure}

The data were reduced using the HARPS Data Reduction Software (DRS) and the RVs were computed using the CCF method (\citealt{2002Msngr.110....9P}, and references therein). We used the same mask and width as those used to extract CORALIE data. The DRS pipeline also provides values for the full width at half-maximum (FWHM) depth of the CCF, the CCF bisector span (BIS), and its equivalent width ($W_{\rm CCF}$, see \citealt{2019MNRAS.487.1082C} for more details). We also extracted the $\log R'_{HK}$ chromospheric index via \cite{1984ApJ...279..763N} from the S-index provided by the pipeline, with bolometric corrections using the $B-V$ colour of the star \citep{1982A&A...107...31M}. Figure \ref{fig:159_gls} reports the spectroscopic time series. It is worth noting that a clear correlation (Pearson's coefficient $=-0.75$) between BIS and RV time series can be seen in the upper right panel.

%%%%%%%%%%%%%%%%%%%%%%%%%%%%%%%%%%%%%%%%%%%%%%%%%%%%%%%%%%%%%%
\section{Stellar parameters}
\label{sec:stellar}

To establish TOI-159 parameters, we produced a co-added spectrum using all the HARPS data. Then, we proceed as follows: 

\begin{itemize}
 \item We used an arbitrary but pre-selected spectrum as a reference, then divided the i$^{th}$ order of each one of other 19 spectra for the i$^{th}$ order of the reference spectrum.
 \item We fitted a third-order polynomial to each spectral ratio and each i$^{th}$ order was then multiplied by this polynomial fit. In this way each order of the 19 spectra is exactly superimposed to the orders of the reference spectrum.
 \item All the orders were then combined separately using a 2-sigma clipping rejection in order to remove possible outliers (cosmic rays and bad pixels).
 \item Each order now has a much greater S/N, and it is easy to apply a normalisation using a high-order polynomial fitting, where spectral lines can be easily identified and ignored.
 \item Finally, the normalised orders were combined in a single spectrum, ready to be used for parameter determination.
\end{itemize}

The combined spectrum has an S/N $\sim$ 150 per extracted pixel at around 6000 Å. 

\subsection{Atmospheric parameters, metallicity and projected rotational velocity}
\label{sec:atm_param}

We used the methodology described in \citet{2019ApJ...882..174V} for the following analysis, where we derived the effective temperature, $T_{\rm eff}$, the surface gravity, $\log{g}$, the microturbulence velocity, $\xi$, the iron abundance, [Fe/H], and the projected rotational velocity, $v \sin{i}_{\star}$, of TOI-159.

For our star, we assumed the initial atmospheric parameters to be $T_{\rm eff}$ = 7000 K, $\log(g) = 4.5$~dex, and $\xi$ = 2.00 km s$^{-1}$. Atmospheric models were calculated using the ATLAS9 code \citep{1970SAOSR.309.....K}, assuming our initial estimations of $T_{\rm eff}$, $\log(g)$, and $\xi$, and [Fe/H]=0. Then $T_{\rm eff}$, $\log(g)$, and $\xi$ were re-adjusted and new atmospheric models calculated in an interactive way in order to remove trends in excitation potential and reduced equivalent width (EW) versus abundance for $T_{\rm eff}$ and $\xi$, respectively, and to satisfy the ionisation equilibrium for $\log(g)$. 83 FeI lines and 7 FeII lines were used for this purpose. Their EW was measured by a gaussian fit, that turned out to map very well the shape of the lines. The [Fe/H] value of the model was changed at each iteration according to the output of the abundance analysis. The local thermodynamic equilibrium programme MOOG \citep{1973ApJ...184..839S} \footnote{https://www-as-utexas-edu.translate.goog/~chris/moog.html} was used for the abundance analysis. NLTE correction for our FeI/II lines \citep{2012MNRAS.427...27B} was also considered during the analysis. These corrections were obtained from the MPIA database \footnote{https://nlte.mpia.de/gui-siuAC\_secE.php}. Projected rotational velocity was obtained by spectrosynthesis on a sample of partially blended FeI lines, assuming an instrumental resolution R=120.000.

Our spectroscopic analysis yields final atmospheric parameters of $T_{\rm eff}$ = 7294 $\pm$ 70 K, log(g) = 4.43 $\pm$ 0.05 dex, and $\xi$ = 2.21 $\pm$ 0.04 km s$^{-1}$. The iron abundance is [Fe/H]$_{NLTE}$= 0.26 $\pm$ 0.04, while $v \sin{i}_{\star}$ turned out to be $22 \pm 2$ km s$^{-1}$. 

\subsection{Independent $v \sin{i}_{\star}$ estimation}
\label{sec:vsini}
We measured the projected rotational velocity $v\sin{i}_{\star}$ using a model-independent Fourier-transform analysis of photospheric absorption lines in the HARPS spectra. The method relies on the fact that rotational broadening produces a characteristic line profile whose Fourier transform exhibits well-defined zeros \citep{1933MNRAS..93..478C}. The position of the first zero is directly proportional to $v\sin{i}_{\star}$ while other broadening mechanisms (e.g., thermal, stark, macroturbulence, or the instrumental profile) do not generate such zeros and therefore do not bias the measurement. We selected a clean set of unblended metal lines across the HARPS spectral range, rejected lines affected by blends, and computed $v\sin{i}_{\star}$ on a line-by-line basis using all 20 available HARPS epochs. The final value and uncertainty were obtained from the distribution of individual measurements, yielding $v\sin{i}_{\star}= 22.6 \pm 4.3~\mathrm{km\,s^{-1}}$, where the quoted uncertainty reflects the line-to-line scatter. This result is fully consistent with the spectroscopic $v\sin{i}_{\star}$ reported above and with the stellar rotation period inferred from photometry, providing independent confirmation of the stellar rotational broadening based purely on the line-profile shape.

\subsection{Infrared flux method and isochrones}
\label{sec:irfm}
We determined the stellar radius and mass of TOI-159 from its atmospheric parameters ($T_{\rm eff}$, $\log{g}$) using a customised version of a code that incorporates the infrared flux method \citep[IRFM,][]{1977MNRAS.180..177B, 2017AJ....153..136S} and isochrone fitting \citep[MCMCI code,][]{2020A&A...635A...6B}. We first computed the stellar angular diameter and $T_{\rm eff}$ using known relations and the apparent bolometric flux. Specifically, the atmospheric parameters are used as priors to form a spectral energy distribution (SED) from stellar atmospheric models \citep[taken from the ATLAS catalogues, ][]{2003IAUS..210P.A20C}, adjusted for extinction estimates \citep{2016ApJ...818..130B, 2019ApJ...887...93G}. The stellar bolometric flux is then derived by comparing synthetic photometry computed from the SED with observational data from the \textit{Gaia} $G$, $G_{\rm BP}$, $G_{\rm RP}$, 2MASS $J$, $H$, $K$, and WISE W1 and W2 bandpasses. We then estimated $R_\star$ by converting the stellar angular diameter using the offset-corrected \textit{Gaia} Early Data Release 3 parallax \citep{2021A&A...649A...4L}. With $R_\star$ as input, we determined the isochronal mass $M_\star$ and age $t_\star$ through the isochrone placement algorithm \citep{2015A&A...575A..18B}, which interpolates the parameters within precomputed grids of PARSEC v1.2S \citep{2017ApJ...835...77M} isochrones and tracks. We report the resulting values in Table~\ref{tab:star_param}.

\subsection{Stellar companion and host star determination}
\label{sec:companion}
TOI-159 has a physical stellar companion (TOI-159~B) that is 3.3 mag fainter and located 0.65$\arcsec$ away (projected distance $\approx$ 225 au), recently discovered by high-resolution imaging observations \citep{2020AJ....159...19Z,2022AJ....164...56L}. Comparing direct imaging observations of TOI-159 taken six years apart \citep{2020AJ....159...19Z,2025FrASS..1208411H} reveals that the photocenter separation between the primary and its bounded companion star remained almost unchanged. In contrast, if the companion were a background object, a clear change in the projected separation would be observable between the two epochs due to different stellar proper motions. By comparing the observed stellar densities of both stars to empirical values \citep[e.g.][]{2003ApJ...585.1038S}, the same authors report that it is uncertain which component is the host of the transiting planet. However, they also underline that if TOI-159~B were the stellar host, this would imply an unrealistically small (large) planetary density (radius). 

The density from the refined stellar parameters of TOI-159 (sect. \ref{sec:irfm}) presented in this work is instead compatible with the density resulting from the refined transit fit (sect. \ref{sec:analysis}), confirming that the primary can therefore be assumed to be the true host star. In contrast, if we assume that the secondary star TOI-159~B is the host, the stellar densities derived from the models and transit fit are no longer compatible.

Given the difference in magnitude between the two binary components, which translates to a small contamination ratio ($CR = F_{\rm secondary} / F_{\rm primary}$) of approximately 0.048, it is also possible to exclude any significant contribution from the secondary star in the spectrophotometric analysis (Sect. \ref{sec:modelling}). Moreover, the magnitude difference likely indicates that TOI-159~B is an early K dwarf. Specifically, we quantified the flux contribution from TOI-159~B (and its wavelength dependence) by dividing two low-resolution model spectra from the \cite{1998PASP..110..863P} library. Assuming F0V and K3V spectral type for the primary and secondary star, respectively, and a difference in absolute magnitude corresponding to the values tabulated by \cite{2013ApJS..208....9P}, we get a contamination factor $CR$ ranging from 0.02 to 0.04 in the IMACS spectral range (4750-6990$\AA$). Even in the most pessimistic assumption that all the contaminating flux would fall into the spectroscopic aperture, that would translate into an overall contamination-induced slope of 2\% at most in our transmission spectrum, much smaller than both our error bars and our best-fit slope (15\%).

\subsection{Chromospheric activity}
\label{sec:chrom}
The mean S-index values calibrated to the Mt. Wilson scale \citep{1995ApJ...438..269B} is 0.23 for TOI-159, which corresponds to $\log R'_{HK}$ values of $-4.55\,\pm\,0.04$ (arithmetic mean and standard deviation). In accordance with the relations from \cite{2018A&A...616A.108B}, the $\log R'_{HK}$ value we obtained indicates a fairly active star.

\subsection{Rotation period}
\label{sec:rot_per}
\begin{figure}
   \centering
   \includegraphics[width=\hsize]%
   {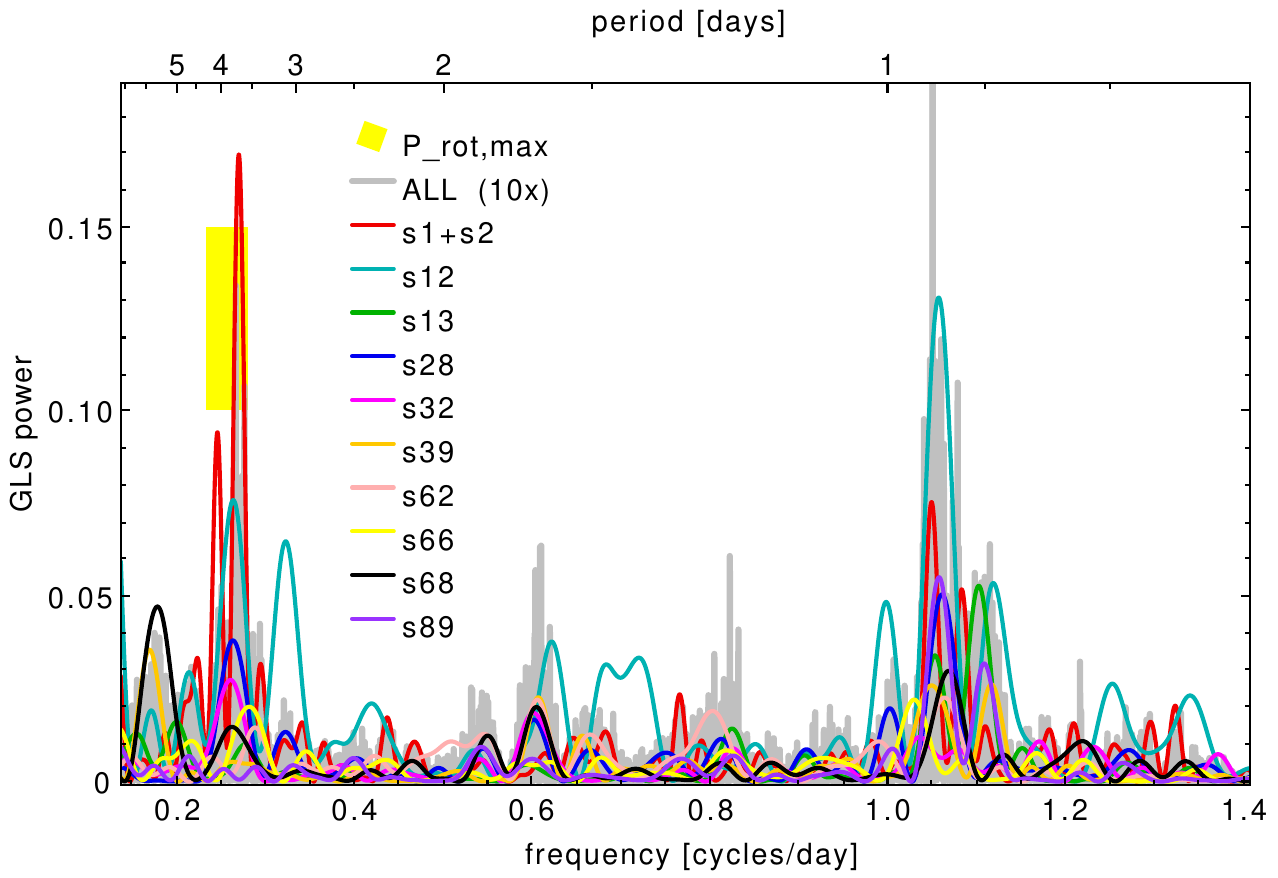}
   \caption{GLS periodogram extracted from \textit{TESS} photometry after modelling TOI-159.01 transits. Both the result for the full light curve (gray thick line) and for just the individual sectors (thin lines colour-coded according to the legend) are plotted. The yellow box shows the expected rotation period 1-$\sigma$ range estimated from spectroscopy (see Sect.~\ref{sec:rot_per} for details). }
   \label{fig:159_gls_tess}
\end{figure}

We can estimate the stellar rotation period from the derived stellar radius and the spectroscopically measured $v \sin{i}_{\star}$. This results in $P_{\rm rot} \sin{i}_{\star}$ = $3.92 \pm 0.36$ d. Subsequently, we attempted to further constrain the rotation period by examining the \textit{TESS} light curves and extracting the generalised Lomb-Scargle (GLS) periodogram \citep{2009A&A...496..577Z} for both the full light curves and for the individual TESS sectors. The GLS periodogram shows three prominent peaks which are not in integer ratios to each other. One at about 4 d, close to the expected $P_{\rm rot}$, one at about 1.7 d and a third one at about 0.95 d (Fig. \ref{fig:159_gls_tess}). The latter two peaks, in particular, are close to the typical $\gamma$ Dor variable stars pulsation range \citep[e.g.][]{1999PASP..111..840K, 2011AJ....142...39H}.

\subsection{Frequency analysis and mode identification}
\label{sec:pulsation}
\begin{figure}
   \centering
   \includegraphics[width=\hsize]%
   {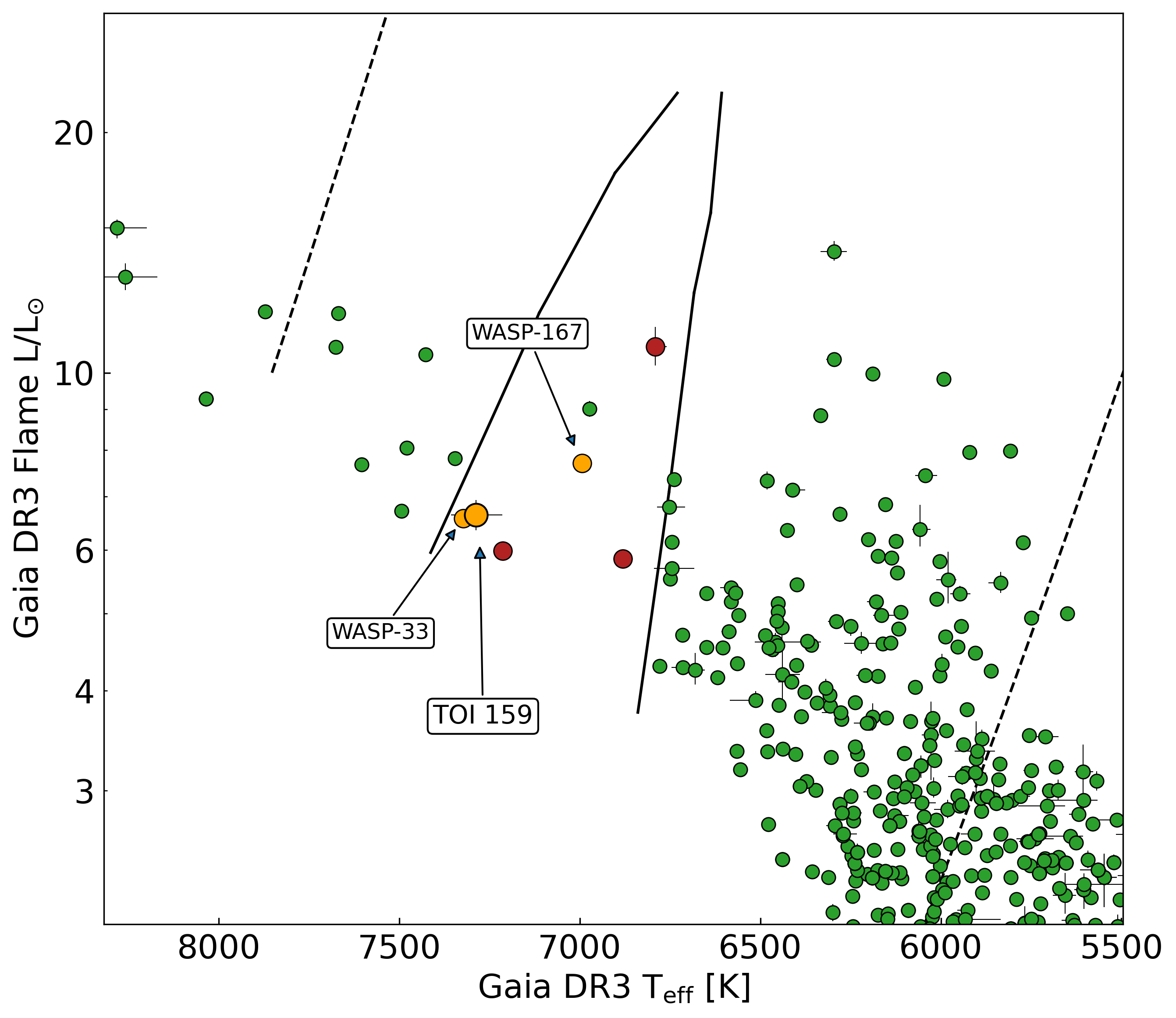}
   \caption{\textit{Gaia} DR3 planet$-$hosting stars on the H$-$R diagram. The black lines show the boundaries of the theoretical instability strip for $\gamma$ Dor stars from \cite{2005A&A...435..927D}. Dashed lines represent the theoretical blue and red edges of the instability strips for g-modes with $l = 2$ from \cite{2016MNRAS.457.3163X}. Orange dots show literature$-$confirmed $\gamma$ Dor stars, while red dots represent stars that exhibit pulsations but have not been classified as $\gamma$ Dor stars. }
   \label{fig:159_hr}
\end{figure}

The stellar parameters of TOI-159 place it in the instability strip, near the $\gamma$ Doradus class (see Fig. \ref{fig:159_hr}). These stars typically have spectral types ranging from A7 to F5, with $T_{\rm eff}$ approximately between 6750 and 7950 K, and show multiple nonradial g-modes pulsations with periods spanning from 8 hours to 5 days \citep{2010aste.book.....A, 2011AJ....142...39H}. To gain a deeper understanding of the pulsation frequencies of TOI-159, we focused on \textit{TESS} sector 32 (i.e. the largest dataset sector with the fewest gaps), with the aim of minimising gaps in the light curve. An important step in this analysis was to model the transits and generate a light curve as if the planet were not transiting. In order to achieve this, we first modelled the full light curve using a Gaussian Process, while simultaneously modelling the transits (see Sect. \ref{sec:analysis}). This approach allowed us to preserve the intrinsic variability of the star when we removed the transit model from the light curve. Finally, we extracted the GLS periodogram and excluded the frequency corresponding to the stellar rotation period and its harmonics.

\begin{figure}
   \centering
   \includegraphics[width=0.8\hsize]%
   {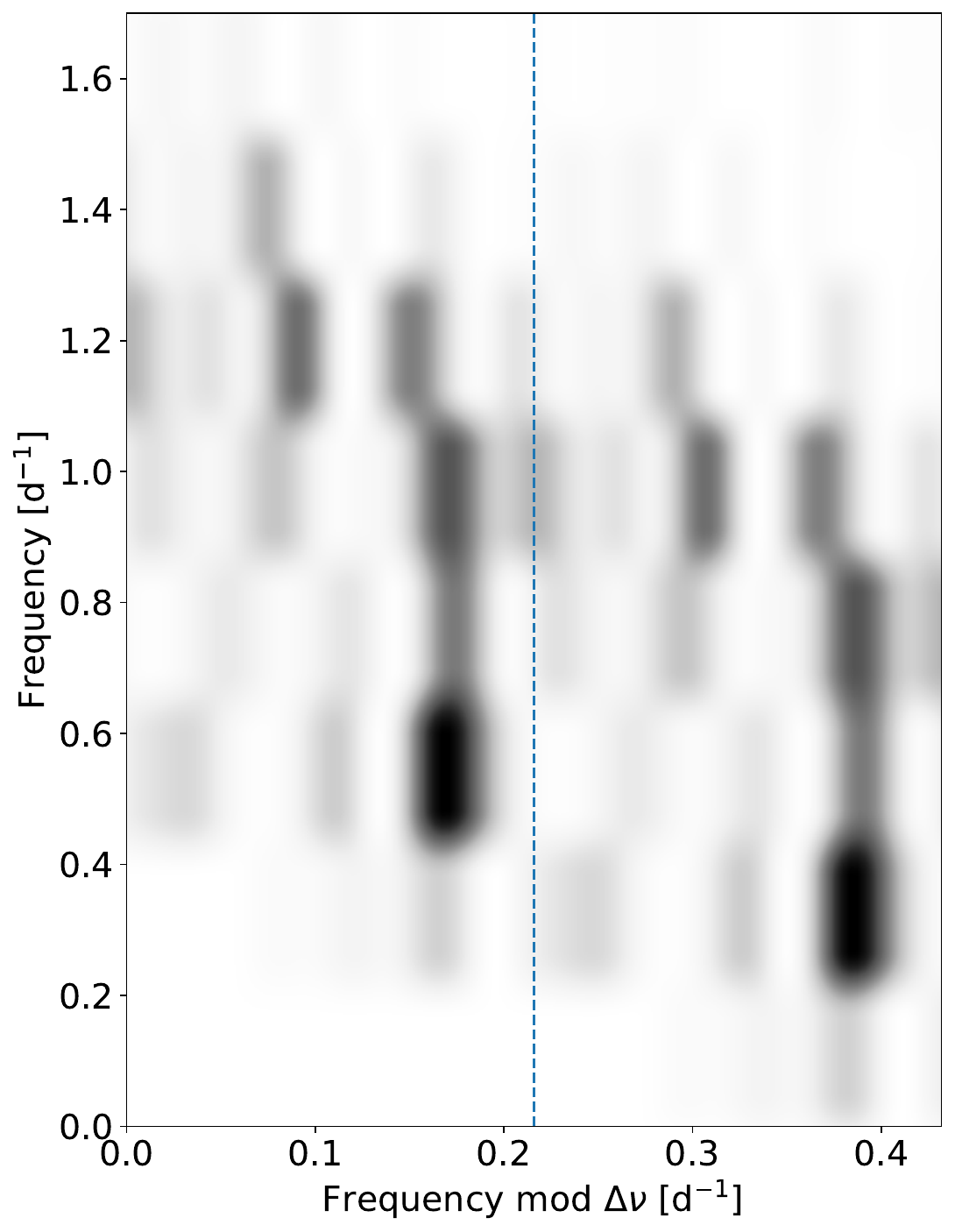}
   \caption{Échelle diagram of TOI-159 sector 32 with a frequency separation $\Delta \nu = 0.216$ d$^{-1}$. A clear ridge can be seen -- probably due to the quadrupole $l = 2$ mode -- while a tentative pattern corresponding to $l = 1$ is emerging. The vertical line corresponds to the frequency spacing. }
   \label{fig:159_echelle}
\end{figure}

\begin{figure}
   \centering
   \includegraphics[width=\hsize]%
   {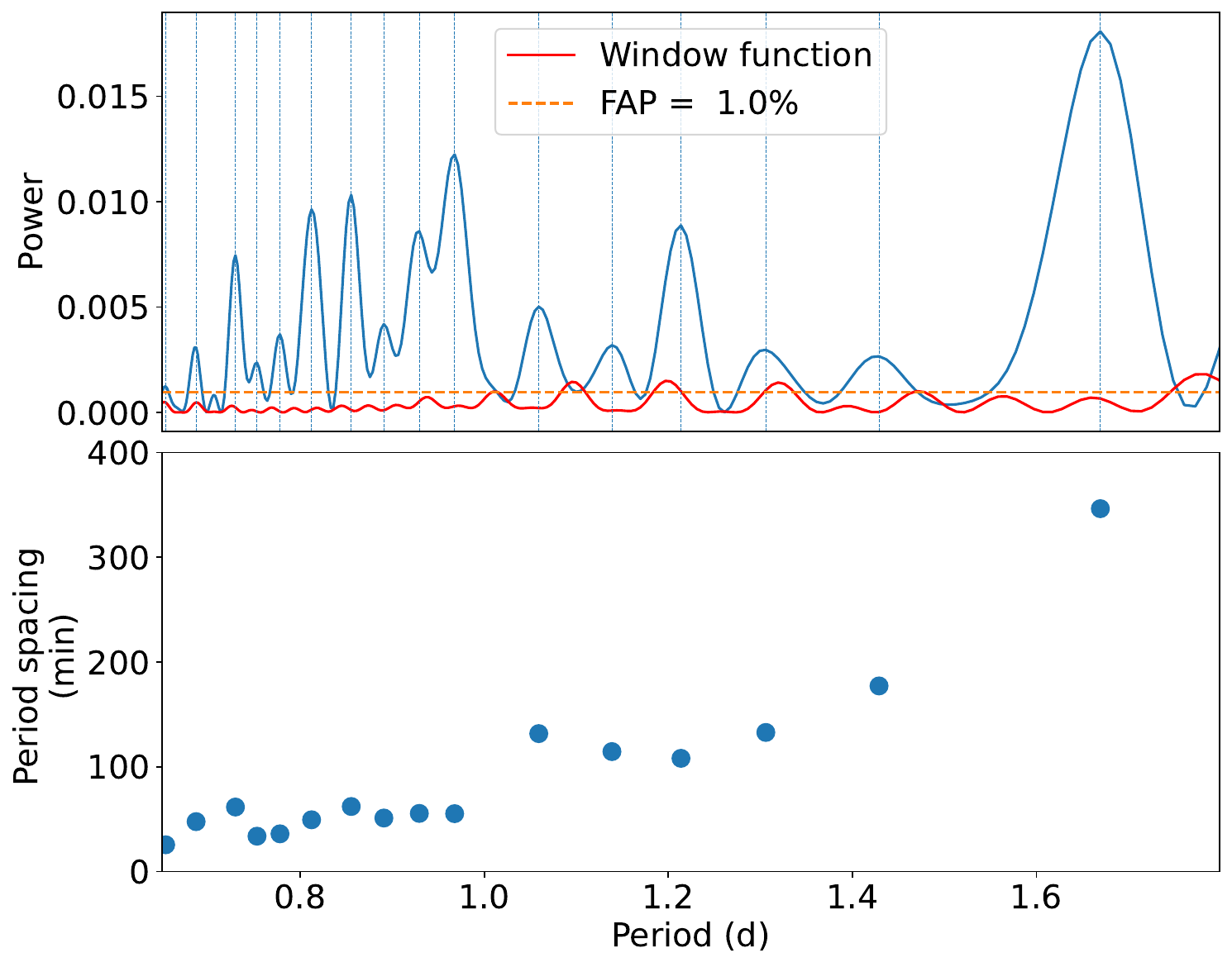}
   \caption{Power spectrum and period spacing diagram of TOI-159. \textit{Top}: GLS power spectrum with x-axis showing the period. Each vertical line corresponds to the location of g modes whose power was greater than a FAP equal to 1\%. \textit{Bottom}: Period spacing diagram showing the spacing between the individual g modes seen in the top panel. }
   \label{fig:159_freq}
\end{figure}

\begin{figure}
   \centering
   \includegraphics[width=\hsize]%
   {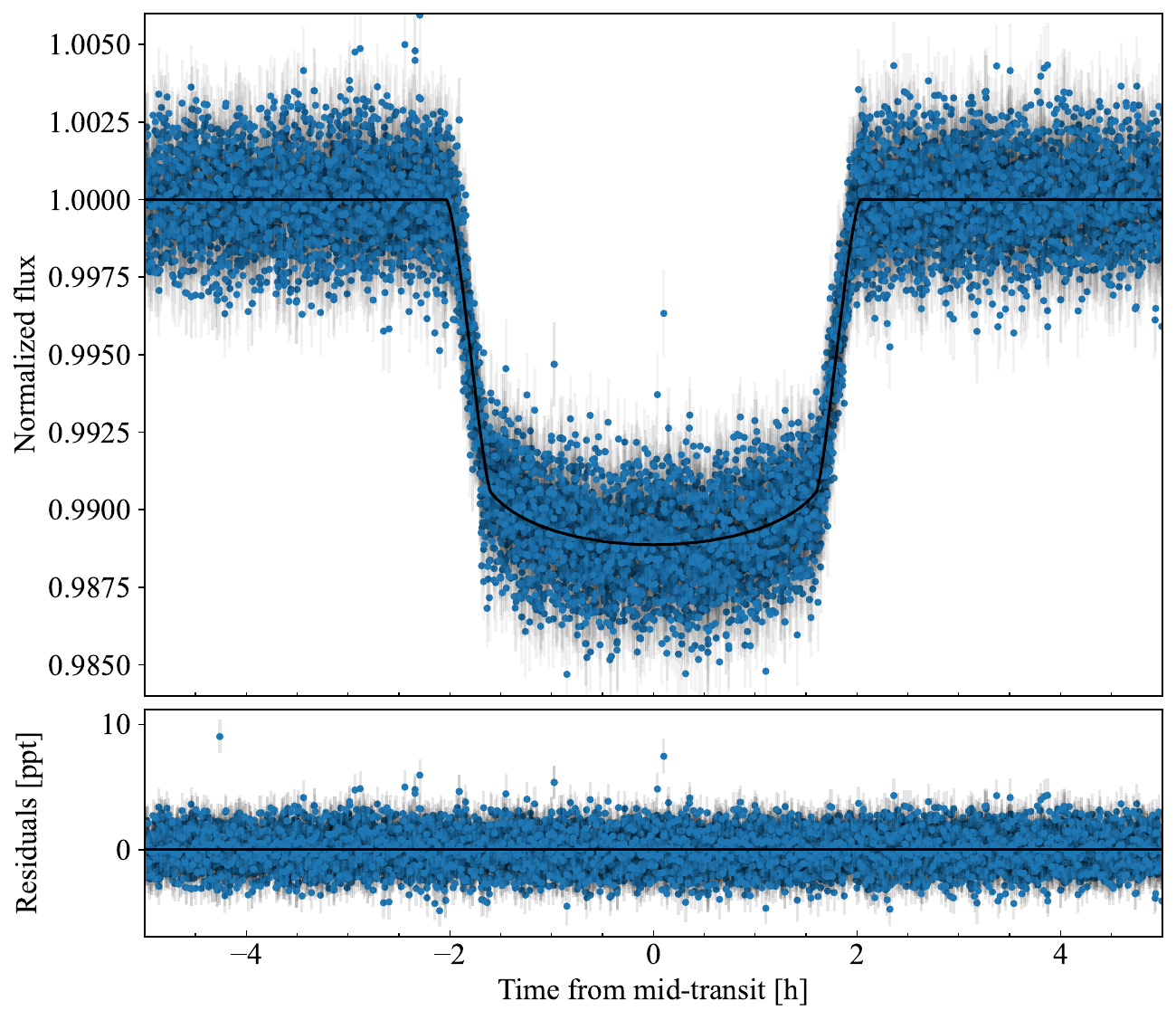}
   \caption{Photometric modelling of TOI-159~b planetary signal. In the \textit{top panel}, we display the \textit{TESS} phase-folded transits of TOI-159~b after normalisation together with the transit model (black line). In the \textit{panel below}, we show the residuals of the joint fit.}
   \label{fig:159lc}
\end{figure}

To identify non-radial g-modes, we inspected the échelle diagram for TOI-159 \citep{2010aste.book.....A, daniel_hey_2020_3629933, 2021A&A...655A..63Y} and the results are shown in Fig. \ref{fig:159_echelle}. A quasi-vertical ridge appears when we use a frequency separation $\Delta \nu$ of about 0.216 d$^{-1}$, which highlights modes with the same angular degree ($l$). Another tentative pattern appears at a lower frequency spacing, which might correspond to a lower angular mode $l$ \citep{2021RvMP...93a5001A}. Moreover, TOI-159 g-mode pulsations are unevenly spaced in period due to the effect of moderate stellar rotation \citep{2017MNRAS.465.2294O}. This can be clearly seen in the period versus period spacing diagram \citep[Fig. \ref{fig:159_freq},][]{2010aste.book.....A,2015ApJS..218...27V, 2015MNRAS.454.1792K, 2021A&A...655A..63Y}. Most importantly, this plot shows a steep upward slope, indicative of retrograde modes that are shifted in the frequency domain due to moderate -- but not rapid -- stellar rotation \citep{2015ApJS..218...27V}. In addition, the steepness of the slope (around 0.16 and Pearson correlation = 0.96) may indicate the presence of a quadrupole mode ($l = 2$) rather than a dipole mode \citep[$l = 1$; see e.g. Fig. 4 of ][]{2017MNRAS.465.2294O}.

In conclusion, TOI-159 appears to be a classical $\gamma$ Dor pulsator with two main non-radial g-modes, as predicted by its $T_{\rm eff}$, $M_\star$ and $L_\star$.

\begin{table}
   \caption[]{Stellar properties of TOI-159.}
     \label{tab:star_param}
     \small
     \centering
       \begin{tabular}{lcc}
         \hline
         \noalign{\smallskip}
         Parameter   &  \object{TOI-159} & Reference  \\
         \noalign{\smallskip}
         \hline
         \noalign{\smallskip}
$\alpha$ (ICRS)                &   ~~03:27:07.683      & {\it Gaia} DR3  \\
$\delta$ (ICRS)                &   $-$78:32:59.420   & {\it Gaia} DR3  \\
$\mu_{\alpha}$ (mas yr$^{-1}$)  &    11.728 $\pm$ 0.015 & {\it Gaia} DR3  \\
$\mu_{\delta}$ (mas yr$^{-1}$)  &    1.872 $\pm$ 0.016  & {\it Gaia} DR3  \\
RV     (km s$^{-1}$)            &    15.03 $\pm$ 0.41   & {\it Gaia} DR3  \\
$\pi$  (mas)                    &    2.869 $\pm$ 0.012  & {\it Gaia} DR3  \\
\noalign{\medskip}
V (mag)                 &  10.930 $\pm$ 0.023   &   \cite{2016yCat.2336....0H}   \\
$B-V$ (mag)             &  0.491 $\pm$ 0.026    &   \cite{2016yCat.2336....0H}  \\
$G$ (mag)               &  10.8561 $\pm$ 0.0046 &   {\it Gaia} DR3  \\
$G_{BP}-G_{RP}$ (mag)   &        0.643 $\pm$ 0.007        &   {\it Gaia} DR3  \\
$J$ (mag)    &   10.038 $\pm$ 0.023 & 2MASS  \\
$H$ (mag)    &   9.835 $\pm$ 0.023 & 2MASS  \\
$K$ (mag)    &   9.758 $\pm$ 0.020 & 2MASS  \\
\noalign{\medskip}
$T_{\rm eff}$ (K)        &  7294 $\pm$ 70 & Spec; Sect. \ref{sec:atm_param} \\  %specanalysis
${\rm [Fe/H]}$ (dex)     &  0.26 $\pm$ 0.04  & Spec; Sect. \ref{sec:atm_param} \\ %HARPS
$\log g$                 &  4.43 $\pm$ 0.05  & Spec; Sect. \ref{sec:atm_param} \\ %specanalysis
$T_{\rm eff}$ (K)        &  7287$^{+69}_{-73}$ & SED; Sect. \ref{sec:irfm} \\  %IRFM+MCMCI
${\rm [Fe/H]}$ (dex)     &  0.30 $\pm$ 0.01  & SED; Sect. \ref{sec:irfm} \\ %IRFM+MCMCI
$\log g$                 &  4.23 $\pm$ 0.01 & SED; Sect. \ref{sec:irfm} \\ %IRFM+MCMCI

\noalign{\medskip}
$\log R^{'}_{\rm HK}$    &     $-$4.55 $\pm$ 0.04 &  Sect. \ref{sec:chrom} \\  %HARPS
$v\sin{i_{\star}}$ (km s$^{-1}$)      &   22 $\pm$ 2 & Sect. \ref{sec:atm_param} \\ 
$v\sin{i_{\star}}$ (km s$^{-1}$)      &   22.6 $\pm$ 4.3 & Sect. \ref{sec:vsini} \\ 
$P_{\rm rot}$ (d)  &    3.95 $\pm$ 0.01 & Sect. \ref{sec:analysis} \\
\noalign{\medskip}
Luminosity (L$_{\odot}$) &    6.64 $\pm$ 0.29 & Sect. \ref{sec:irfm} \\
Radius (R$_{\odot}$)     &    1.620 $\pm$ 0.014 & Sect. \ref{sec:irfm} \\
Mass (M$_{\odot}$)       &    1.661$^{+0.017}_{-0.019}$ & Sect. \ref{sec:irfm} \\
Density ($\rho_{\odot}$)       &    0.39 $\pm$ 0.01 & Sect. \ref{sec:irfm} \\
Age  (Myr)               &    145$^{+31}_{-23}$ & Sect. \ref{sec:irfm} \\
Distance  (pc)           &   347.2$^{+1.2}_{-1.4}$ & \cite{2021AJ....161..147B} \\

         \noalign{\smallskip}
         \hline
      \end{tabular}
\end{table}

%%%%%%%%%%%%%%%%%%%%%%%%%%%%%%%%%%%%%%%%%%%%%%%%%%%%%%%%%%%%%%
\section{Analysis}
\label{sec:analysis}
\label{phot}

\subsection{Probabilistic validation and blends exclusion}
Because of the low angular resolution of the \textit{TESS} cameras, a fraction of the objects initially identified as candidate exoplanets are expected to be false positives (FPs). Therefore, we conducted a probabilistic validation procedure, which aims at distinguishing between a planet and a FP from a specific transiting candidate \citep[e.g.,][]{2011ApJ...727...24T,2012ApJ...761....6M,2014MNRAS.441..983D}. As a first step, we checked \textit{Gaia} DR3 photometry to examine the area corresponding to the \textit{TESS} PSF size around the host star to exclude each \textit{Gaia} source as possible blended eclipsing binaries mimicking the transit signal (see Fig. \ref{fig:field}). As further evidence for the planetary transit source origin, we performed centroid motion tests \citep{2020MNRAS.498.1726M,2020MNRAS.495.4924N}. Moreover, to rigorously quantify the probability of each candidate planet to be a FP, we utilised the \texttt{VESPA}\footnote{\url{https://github.com/timothydmorton/VESPA}} software \citep{2012ApJ...761....6M} by following the procedure adopted in \cite{2022MNRAS.516.4432M}, which consider the main issues reported in \cite{2023RNAAS...7..107M} and allows us to obtain reliable results when using \texttt{VESPA}. TOI-159.01 has an extremely low FP probability (FPP $< 10^{-6}$), which statistically validates this planet. As a matter of precaution, we repeated this calculation using \texttt{TRICERATOPS} \citep{2021AJ....161...24G}. We found that FPP $= 0.0001 \pm 0.0002$ and nearby FP probability NFPP $= 0$. We also included contrast curves from published high-resolution imaging data \citep{2020AJ....159...19Z, 2022AJ....164...56L} and found that the updated FPP is $0.0008 \pm 0.0024$. These results support our previous findings. This validation has been instrumental to the follow-up observations presented in Sect. \ref{sec:harps} and \ref{sec:modelling}, which definitively confirm the planetary nature of what we will henceforth call TOI-159~b. 

To exclude the possibility of a blended false positive on the secondary star contaminating the RV measurements \citep[e.g.][]{2005ApJ...621.1061M, 2025PSJ.....6..300W}, we inspected whether the BIS changes as a function of the stellar rotation period rather than the orbital period. We ran the GLS periodogram for the BIS and identified the strongest peak at 3.95 days. This is entirely consistent with the stellar rotation period value found with our joint photometry and RV analysis (see Table \ref{tab:star_param}). Subsequently, we removed the signal associated to the strongest peak and ran the GLS again, finding no signals compatible with the 3.7-d orbital period. This analysis therefore rules out the scenario of a blended tertiary star with the broad lines of the primary. 

\subsection{Combined HARPS, TESS, and IMACS modelling}
\label{sec:modelling}
To estimate the planetary properties of TOI-159~b, we simultaneously studied all \textit{TESS} photometry, IMACS spectrophotometry, and HARPS RV time series. This analysis has been performed in a Bayesian framework using \texttt{PyORBIT}\footnote{\url{https://github.com/LucaMalavolta/PyORBIT}} \citep{2016A&A...588A.118M,2018AJ....155..107M}, a Python package that models planetary transits and RVs while simultaneously considering stellar activity effects. For this paper, we tried various approaches to model the activity using Gaussian processes \citep[GPs,][]{rasmussen2006gaussian, 2014MNRAS.443.2517H} and harmonic models.

We simultaneously modelled the \textit{TESS} (Section~\ref{sec:tess}) and IMACS (Section~\ref{sec:imacs}) transit light curves with the \texttt{batman} code \citep{2015PASP..127.1161K}, the stellar activity and the planetary signal in the RV series, by fitting the following parameters: the reference transit time ($T_0$), the planetary-to-star radius ratio ($R_p /R_{\star}$), the impact parameter ($b$), the orbital period ($P_{\rm orb}$), the stellar density ($\rho_\star$), the RV semi-amplitude ($K$), the quadratic limb-darkening (LD) coefficients, $ u_1$ and $u_2$, adopting the LD parameterisation introduced by \cite{2013MNRAS.435.2152K}, the systemic RV (offset), and a jitter term added in quadrature to the photometric and RV errors to account for effects not captured by our model (e.g. short-term stellar activity) or any potential underestimation of the error bars. We fitted the orbital periods and semi-amplitudes of the RV signal in linear space. We imposed Gaussian priors on the LD coefficients $u_1$ and $u_2$. We estimated the LD coefficients using \texttt{PyLDTk}\footnote{\url{https://github.com/hpparvi/ldtk}} \citep{2013A&A...553A...6H,2015MNRAS.453.3821P} and applying the specific filters or wavelength bins used during the observations. We added $0.1$ in quadrature to their associated Gaussian error to account for the known underestimation by models. Moreover, we calculated the eccentricity, $e$, and periastron argument, $\omega$, by fitting $\sqrt{e}\cos{\omega}$ and $\sqrt{e}\sin{\omega}$ \citep{2013PASP..125...83E}. Finally, we imposed a Gaussian prior on the stellar density and uniform priors on $P_{\rm orb}$ and $T_0$. The list of priors can be found in Table \ref{table:model-lcrv}. It is worth noting that, when modelling the IMACS transits, we left $R_p /R_{\star}$ free to vary independently for each channel to generate the transmission spectrum. We would like to emphasise that $R_p$ has been assumed to be the same across all wavelengths; specifically, it has been estimated from the \textit{TESS} data. Spectrophotometry has only been used to refine ephemeris.

We modelled the stellar activity in the RV series using a GP regression (stellar rotation) combined with a simple harmonic model (stellar pulsation). We used a quasiperiodic kernel. As part of the stellar rotation modelling, we set the rotation period $P_{\rm rot}$ (Gaussian prior, as the one defined in Sect. \ref{sec:rot_per}), the characteristic decay timescale $P_{\rm dec}$, and the coherence scale $\omega$. Instead, we set a Gaussian prior for the period of the main pulsation mode $P$, and include two harmonics (one sine, one cosine). We fit the period and harmonic amplitudes in linear space. We modelled the stellar activity in all IMACS transits including a GP with a Matern-3/2 kernel \citep[e.g.,][]{2020A&A...643A..25P}, while we used a \texttt{celerite2} \citep{celerite1, celerite2} rotation-oscillation kernel to model the activity in all \textit{TESS} sectors. As part of this modelling, we set the rotation period $P_{\rm rot}$ as a unique parameter that is shared between the quasiperiodic and the \texttt{celerite2} GPs. The other GP hyperparameters are independent between the photometric and spectroscopic datasets.

\begin{figure}
   \centering
   \includegraphics[width=\hsize]%
   {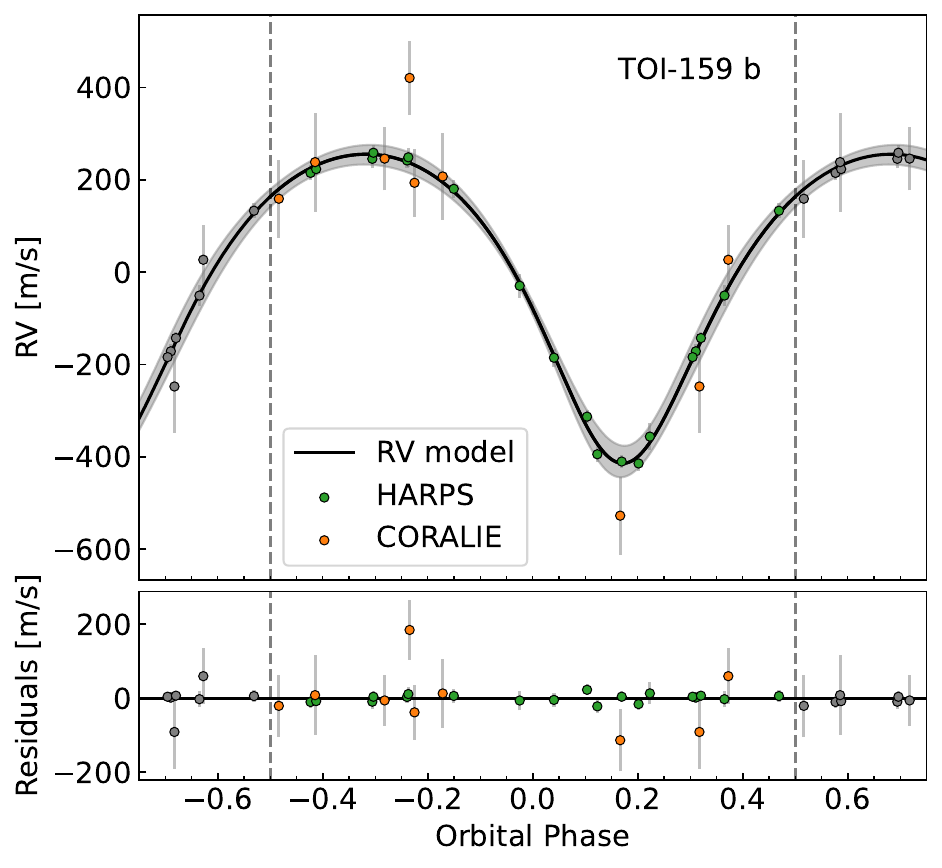}
   \caption{Phase-folded RV curve of TOI-159~b. The shaded area shows the uncertainties ($\pm 1\sigma$) of the RV model. The residuals of the fit are shown in the \textit{bottom} panel.}
   \label{fig:159rv}
\end{figure}

\begin{figure}
   \centering
   \includegraphics[width=\hsize]%
   {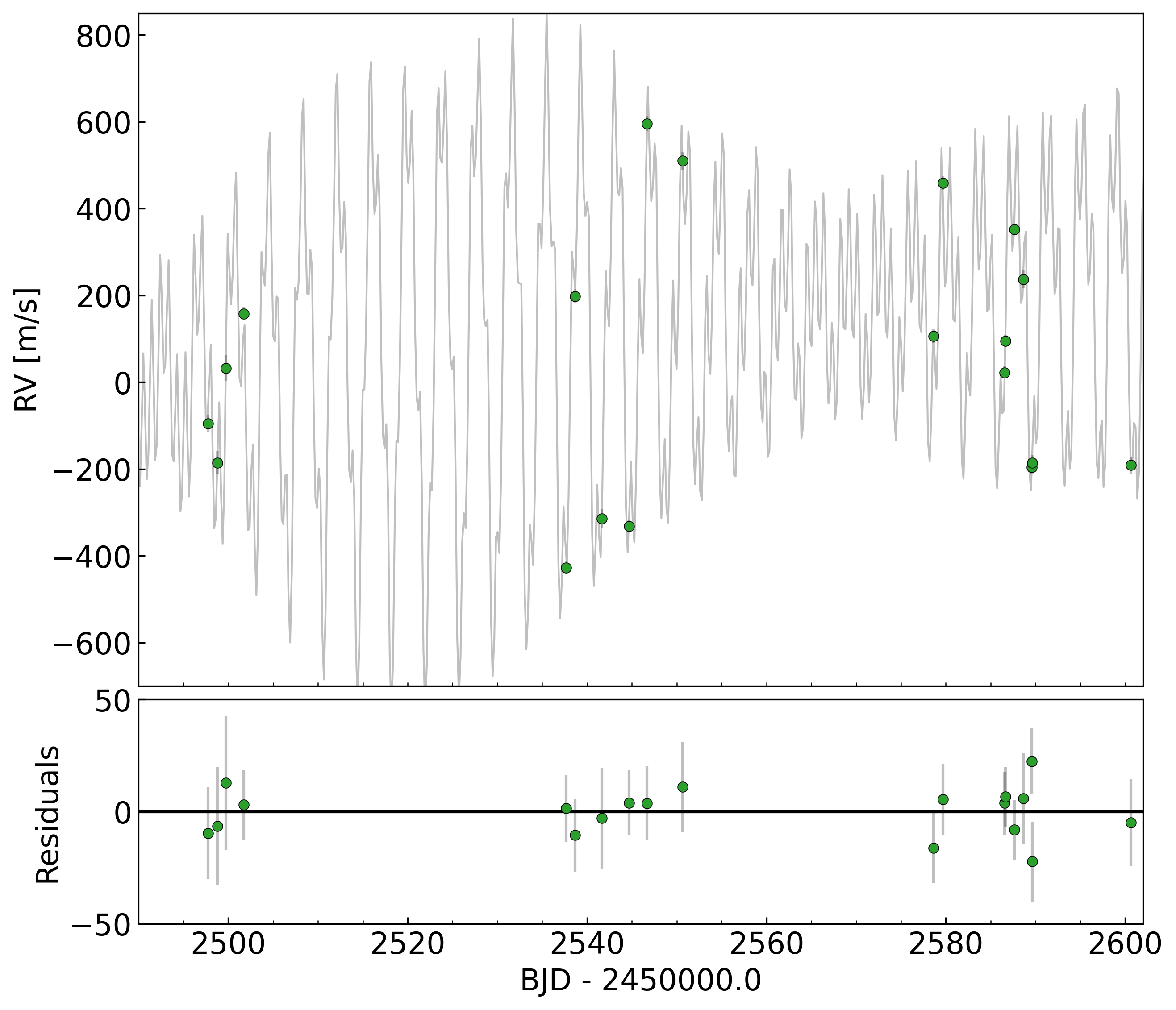}
   \caption{Modelling of HARPS RV time series. \textit{Top:} RV time series with superimposed full Keplerian $+$ stellar activity model. \textit{Bottom:} Residuals of the fit.}
   \label{fig:159rv_full}
\end{figure}

We first performed a global optimisation of the parameters by running a differential evolution algorithm (\citealt{1997JGOpt..11..341S}, \texttt{PyDE}\footnote{\url{https://github.com/hpparvi/PyDE}}) and then a Bayesian analysis. We used the affine-invariant ensemble sampler \citep{2010CAMCS...5...65G} for Markov chain Monte Carlo, as implemented in \texttt{emcee} \citep{2013PASP..125..306F}. We used $4 n_{\rm dim}$ walkers (with $n_{\rm dim}$ representing the model dimensionality) for 100 000 generations with \texttt{PyDE}, followed by 250 000 steps with \texttt{emcee}. We applied a thinning factor of 200 to mitigate the effect of chain auto-correlation. We discarded the first 25 000 steps (burn-in) after checking the convergence of the chains using the Gelman–Rubin statistic \citep{1992StaSc...7..457G}. Figures \ref{fig:159rv}, \ref{fig:159rv_full} and Table \ref{table:model-lcrv} show the result of the modelling.

\begin{table}
\centering
{\small
\caption{Priors and outcomes of joint modelling.}             
\label{table:model-lcrv}         
\addtolength{\tabcolsep}{-0.4em}
\begin{tabular}{p{3.35cm} c c c }     % 7 columns 
\hline\hline     
\multicolumn{2}{c}{Stellar parameters} & \multicolumn{2}{c}{TOI-159} \rule{0pt}{2.0ex} \rule[-1ex]{0pt}{0pt}\\ 
\hline    
\multicolumn{1}{l}{Parameter} & Unit & Prior & Value \rule{0pt}{2.2ex} \rule[-1ex]{0pt}{0pt}\\ 
\hline 
   Density ($\rho_{\star}$) & $\rho_{\sun}$ & $\mathcal{N}$(0.39, 0.01) & 0.389$\pm$0.01   
    \rule{0pt}{2.0ex} \rule[-1ex]{0pt}{0pt}\\
   TESS LD coeff. ($u_1$) & &$\mathcal{N}$(0.31, 0.10) & 0.26$^{+0.03}_{-0.03}$ \rule{0pt}{2.0ex} \rule[-1ex]{0pt}{0pt}\\
   TESS LD coeff. ($u_2$) & &$\mathcal{N}$(0.15, 0.10) & 0.10$^{+0.05}_{-0.05}$ \rule{0pt}{2.0ex} \rule[-1ex]{0pt}{0pt}\\
   TESS sc jitter & ppt & ...  & 0.009$^{+0.010}_{-0.007}$ \rule{0pt}{1.8ex} \rule[-1ex]{0pt}{0pt}\\
   TESS lc jitter & ppt & ...  & 0.07$^{+0.03}_{-0.04}$\rule{0pt}{0.8ex} \rule[-1ex]{0pt}{0pt}\\
   RV jitter (HARPS) & m s$^{-1}$ & ... & 18$^{+24}_{-12}$ \rule{0pt}{2.0ex} \rule[-1ex]{0pt}{0pt}\\
   RV jitter (CORALIE) & m s$^{-1}$ & ... & 109$^{+118}_{-76}$ \rule{0pt}{2.0ex} \rule[-1ex]{0pt}{0pt}\\
   RV offset (HARPS) & m s$^{-1}$ & ... & 13515$^{+278}_{-424}$ \rule{0pt}{1.0ex}\rule[-1ex]{0pt}{0pt}\\
   RV offset (CORALIE) & m s$^{-1}$ & ... & 13544$^{+262}_{-413}$ \rule{0pt}{1.0ex}\rule[-1ex]{0pt}{0pt}\\
   \cline{0-1} 
   \textit{GP parameters} & & & \rule{0pt}{2.2ex} \rule[-0.8ex]{0pt}{0pt}\\
   $H_{\rm amp}$ (HARPS) & m s$^{-1}$ & $\mathcal{U}$(0.1, 2000.0) & 524$^{+259}_{-194}$ \rule{0pt}{2.2ex} \rule[-0.8ex]{0pt}{0pt}\\ %rv
   $H_{\rm amp}$ (CORALIE) & m s$^{-1}$ & $\mathcal{U}$(0.1, 2000.0) & 484$^{+286}_{-228}$ \rule{0pt}{2.2ex} \rule[-0.8ex]{0pt}{0pt}\\
   \cline{0-1} 
    \textit{Harmonics parameters} & & & \rule{0pt}{2.2ex} \rule[-0.8ex]{0pt}{0pt}\\
   S2$_{\rm amp}$ & m s$^{-1}$ & $\mathcal{U}$(0.0, 250.0) & 23$^{+18}_{-13}$ \rule{0pt}{2.2ex} \rule[-0.8ex]{0pt}{0pt}\\ %sine
   C1$_{\rm amp}$ & m s$^{-1}$ & $\mathcal{U}$(0.0, 250.0) & 174$^{+17}_{-20}$ \rule{0pt}{2.2ex} \rule[-0.8ex]{0pt}{0pt}\\ %cosine
\hline     
\hline   
\multicolumn{4}{c}{Stellar activity (GP + Harmonics)} \rule{0pt}{2.3ex} \rule[-0.9ex]{0pt}{0pt}\\ 
\hline    
Parameter & Unit & Prior & Value \rule{0pt}{2.2ex} \rule[-0.9ex]{0pt}{0pt}\\ 
\hline 
   Rotational period ($P_{\rm rot}$) & days & $\mathcal{N}$(3.92, 0.36) & 3.95$\pm$0.02 \rule{0pt}{2.2ex} \rule[-0.9ex]{0pt}{0pt}\\
   Decay Timescale ($P_{\rm dec}$) & days & $\mathcal{U}$(4.70, 1000.0) & 147$^{+102}_{-56}$ \rule{0pt}{2.2ex} \rule[-0.9ex]{0pt}{0pt}\\
   Coherence scale ($w$) &  & $\mathcal{U}$(0.001, 1.50) & 1.06$^{+0.29}_{-0.35}$ \rule{0pt}{2.2ex} \rule[-0.9ex]{0pt}{0pt}\\
   \cline{0-1}
   Pulsation period ($P$) & days & $\mathcal{N}$(0.92, 0.1) & 0.9366$^{+0.0005}_{-0.0006}$ \rule{0pt}{2.2ex} \rule[-0.9ex]{0pt}{0pt}\\
   Phase ($\Theta$) & deg & ... & 301$^{+13}_{-16}$ \rule{0pt}{2.2ex} \rule[-1ex]{0pt}{0pt}\\
\hline
\hline   
\multicolumn{2}{c}{Planet} & \multicolumn{2}{c}{TOI-159~b} \rule{0pt}{2.0ex} \rule[-1ex]{0pt}{0pt}\\ 
\hline    
   Orbital period ($P_{\rm orb}$) & days & $\mathcal{U}$(3.762, 3.764) & 3.7628396  \rule{0pt}{2.2ex} \rule[-1ex]{0pt}{0pt}\\
   & &   & $\pm$0.0000003  \rule[-1ex]{0pt}{0pt}\\
   Central time of transit ($T_{\rm 0}$) & BTJD & $\mathcal{U}$(2382.232, & 2382.23258  \rule[-1ex]{0pt}{0pt}\\
   & &   2382.233) & $\pm$0.00009  \rule[-1ex]{0pt}{0pt}\\
   Scaled semi-maj. axis ($\frac{a}{R_{\star}}$) & & ... & 7.43$\pm$0.06  \rule[-1ex]{0pt}{0pt}\\
   Orbital semi-maj. axis ($a$) & AU & ... & 0.0560(7) \rule{0pt}{2.0ex} \rule[-1ex]{0pt}{0pt}\\
   Orbital inclination ($i$) & deg & ... & 88.50$^{+0.60}_{-0.40}$ \rule{0pt}{2.0ex} \rule[-1ex]{0pt}{0pt}\\
   Orbital eccentricity ($e$) &  & $\mathcal{U}$(0, 0.9) & 0.24$^{+0.03}_{-0.04}$ \rule{0pt}{2.0ex} \rule[-1ex]{0pt}{0pt}\\
   Impact parameter ($b$) & & $\mathcal{U}$(0, 1) & 0.18$^{+0.05}_{-0.07}$ \rule{0pt}{1.0ex} \rule[-1ex]{0pt}{0pt}\\
   Planet/star rad. ratio ($\frac{R_{\rm p}}{R_{\star}}$) & & $\mathcal{U}$(0, 0.5) & 0.1028(3) \rule[-1ex]{0pt}{0pt}\\
   Periastron argument ($\omega$) & deg & ... & $-$176$^{+4}_{-5}$  \rule[-1ex]{0pt}{0pt}\\
   Mean longitude ($L$) & deg & ... & 201$\pm$4 \rule{0pt}{2.0ex} \rule[-1ex]{0pt}{0pt}\\
   Transit duration ($T_{14}$)\tablefootmark{a} & days & ... & 0.176$\pm$0.002   \rule[-1ex]{0pt}{0pt}\\
   RV semi-amplitude ($K$) & m s$^{-1}$ & $\mathcal{U}$(0.01, 2000) & 335$^{+23}_{-26}$ \rule{0pt}{2.0ex} \rule[-1ex]{0pt}{0pt}\\
   Planetary radius ($R_{\rm p}$) & $R_{\oplus}$ & ... & 18.18$\pm$0.16  \rule[-1ex]{0pt}{0pt}\\
    & $R_{\rm J}$ & ... & 1.622$\pm$0.015  \rule[-1ex]{0pt}{0pt}\\
   Planetary mass ($M_{\rm p}$) & $M_{\oplus}$ & ... & 1110$^{+79}_{-84}$   \rule[-1ex]{0pt}{0pt}\\
    & $M_{\rm J}$ & ... & 3.49$\pm$0.26   \rule[-1ex]{0pt}{0pt}\\
   Planetary density ($\rho_{\rm p}$) & g cm$^{-3}$ & ... & 1.02 $\pm$ 0.08 \rule{0pt}{2.0ex}\rule[-1ex]{0pt}{0pt}\\
\hline   
\end{tabular}
 }
 \tablefoot{\tablefoottext{a}{From \cite{2010exop.book...55W}.}}
\end{table}

\subsubsection{Best model solution}
As previously reported, our best model solution is the one in which we let the eccentricity $e$ free to vary. Our choice is motivated by the comparison -- using the Bayesian information criterion \citep[BIC,][]{1978AnSta...6..461S} -- of the Keplerian solution and a second analysis in which we assumed a circular orbit. This comparison showed a strong preference for case 1 (eccentric) over case 2 (circular), with a $\Delta {\rm BIC}_{21}$ of 6.2 \citep{doi:10.1080/01621459.1995.10476572}.

We tried three different approaches to model the stellar activity (rotation and $\gamma$ Dor pulsations) while simultaneously modelling the planetary signal. In particular, we modelled the activity in the RVs with (A) a simple GP, (B) a simple harmonic model, (C) a combination of a GP and a harmonic model. Again, we compared the three cases using the BIC and report the results are the following: BIC$_\textrm{A} = 290.5$, BIC$_\textrm{B} = 400.9$, and BIC$_\textrm{C} = 272.7$. The combined case (C) is strongly favoured over the other two, and it is the one we have taken as reference.

\begin{figure}
   \centering
   \includegraphics[width=\hsize]%
   {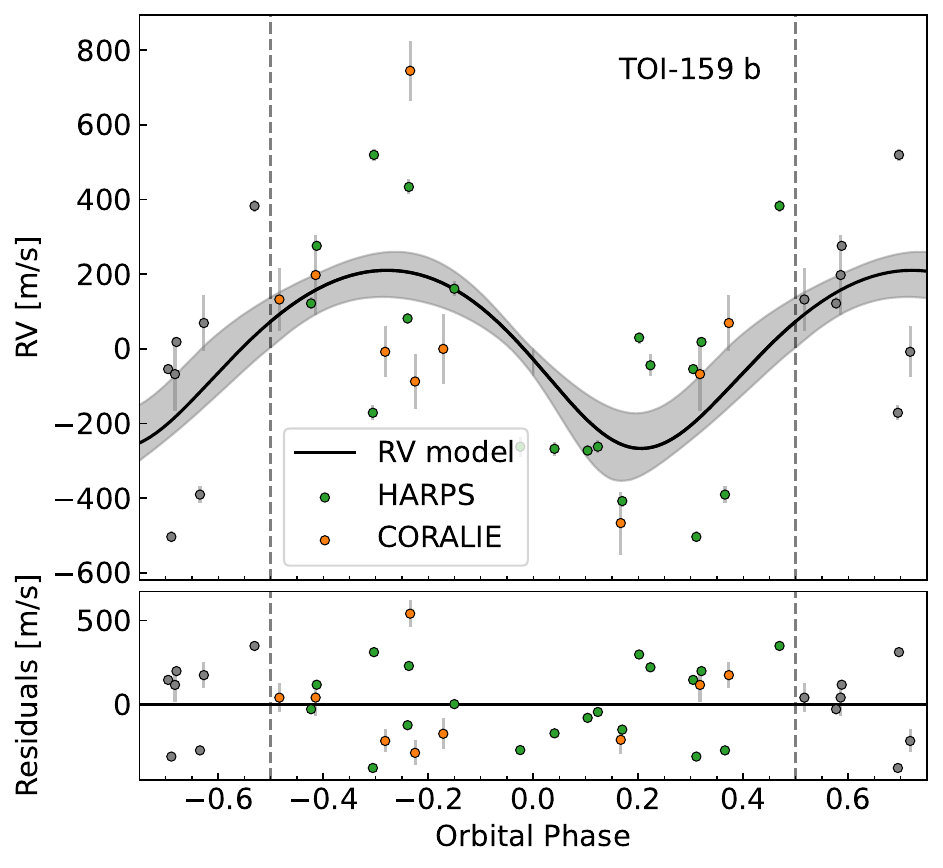}
   \caption{Phase-folded RV curve of TOI-159~b, with no modelling of stellar activity. The residuals of the fit are shown in the \textit{bottom} panel.}
   \label{fig:159rv_noact}
\end{figure}

We also examined the case of no planet signal (case 2) in the RVs and compared it to the case where TOI-159~b signal is included (case 1). The latter is strongly favoured ($\Delta {\rm BIC}_{21} =$ 42), so there is no doubt about the existence of TOI-159~b. Moreover, to emphasise the impact of stellar activity, Fig. \ref{fig:159rv_noact} shows the phase-folded RV for which the stellar activity was not modelled. We would like to emphasise that the typical uncertainty associated with BIC values ranges between 5 and 10, which is much smaller than the difference in BIC values between the models we compared. To estimate these uncertainties, we considered all the parameter sets from the MCMC posteriors and calculated the BIC for each one. 

To test whether the inferred uncertainty on $K$ and $e$ is robust or underestimated when using GP regression \citep[e.g.][]{2023AJ....166...62B, 2024MNRAS.531.4275B}, we performed two different analyses. First, we ran a transit fit alone and retrieved an eccentricity of $e = 0.1_{-0.05}^{ +0.17}$. This first test reinforces the results of our best model solution, favouring an eccentric orbit even without including RVs. We emphasise that transit photometry alone is expected to provide only weak constraints on eccentricity due to strong degeneracies with other transit parameters \citep[e.g.][]{2012MNRAS.421.1166K}. Second, we performed a cross-validation test to evaluate how the retrieved model matches the non-fitted data. Specifically, we repeated our reference model analysis after randomly removing about 25\% of the HARPS and CORALIE data. We found an eccentricity of $e = 0.176_{-0.071}^{+0.055}$ ($2.48\sigma$) and a semi-amplitude of $K = 376_{-55}^{+34}$ m s$^{-1}$. These values are slightly less precise, but are compatible with the reference model and therefore support our joint analysis. In the latter test, the eccentricity is significantly non-zero even after accounting for the Lucy-Sweeney bias \citep{2013A&A...551A..47L}.

\subsection{Planet equilibrium temperature}
\label{sec:tsm}
We computed TOI-159~b equilibrium temperature assuming full day-night heat redistribution and zero albedo with the simple equation: $T_{\rm eq} = T_{\rm eff} \sqrt{\frac{R_\star}{a}}\left(\frac{1}{4}\right)^{1/4}$, with $T_{\rm eff}$ the stellar effective temperature, $R_\star$ the stellar radius, and $a$ the orbital semi-major axis. We found $T_{\rm eq} = 1891 \pm 23$ K. We then inserted the calculated equilibrium temperature into the TSM equation, as described in \cite{2018PASP..130k4401K}. We found TSM $= 44 \pm 4$. This estimated value -- compared to the one predicted by the \textit{TESS} atmospheric characterisation working group (TSM = 121) without knowing the real planet mass -- is lower and does not reach the cut-off for follow-up efforts suggested by \cite{2018PASP..130k4401K}.

\subsection{Transmission spectrum}
A direct output of the global fit described in Section~\ref{sec:modelling}, where an independent $R_p/R_\star$ fit parameter was assigned to each chromatic light curve from IMACS, is the planetary transmission spectrum, i.e., a measurement of $R_p/R_\star$ (or, more frequently, of transit depth $(R_p/R_\star)^2$) as a function of wavelength. Our transmission spectrum of TOI-159~b is plotted in Fig.~\ref{fig:transmission_spectrum}, while the best-fit values of $R_p/R_\star(\lambda)$ are reported in Table~\ref{tab:trans} along with their 1-$\sigma$ uncertainties.

Upon first visual inspection there appears to be possible wavelength-dependent modulation within the spectrum, although with the caveat of the low number of datapoints and their significant error bars. Potential features including a positive bluewards slope and/or a broad absorption feature at $\sim$0.6 $\mu$m warranted further exploration with inverse modelling techniques. If genuine, these features could result from absorption from a planetary atmosphere, contamination from stellar activity (wavelength dependent contrast between the stellar photosphere and the unocculted starspots), or both. A detailed modelling of the transmission spectrum, accounting for both effects, is described in Sect.~\ref{sec:atmoretrieval}.

%%%%%%%%%%%%%%%%%%%%%%%%%%%%%%%%%%%%%%%%%%%%%%%%%%%%%%%%%%%%%%
\section{Discussion}
\label{sec:discussion}

\subsection{TOI-159\,b is an inflated HJ orbiting a $\gamma$ Dor pulsator}
\begin{figure}
   \centering
   \includegraphics[width=\hsize]%
   {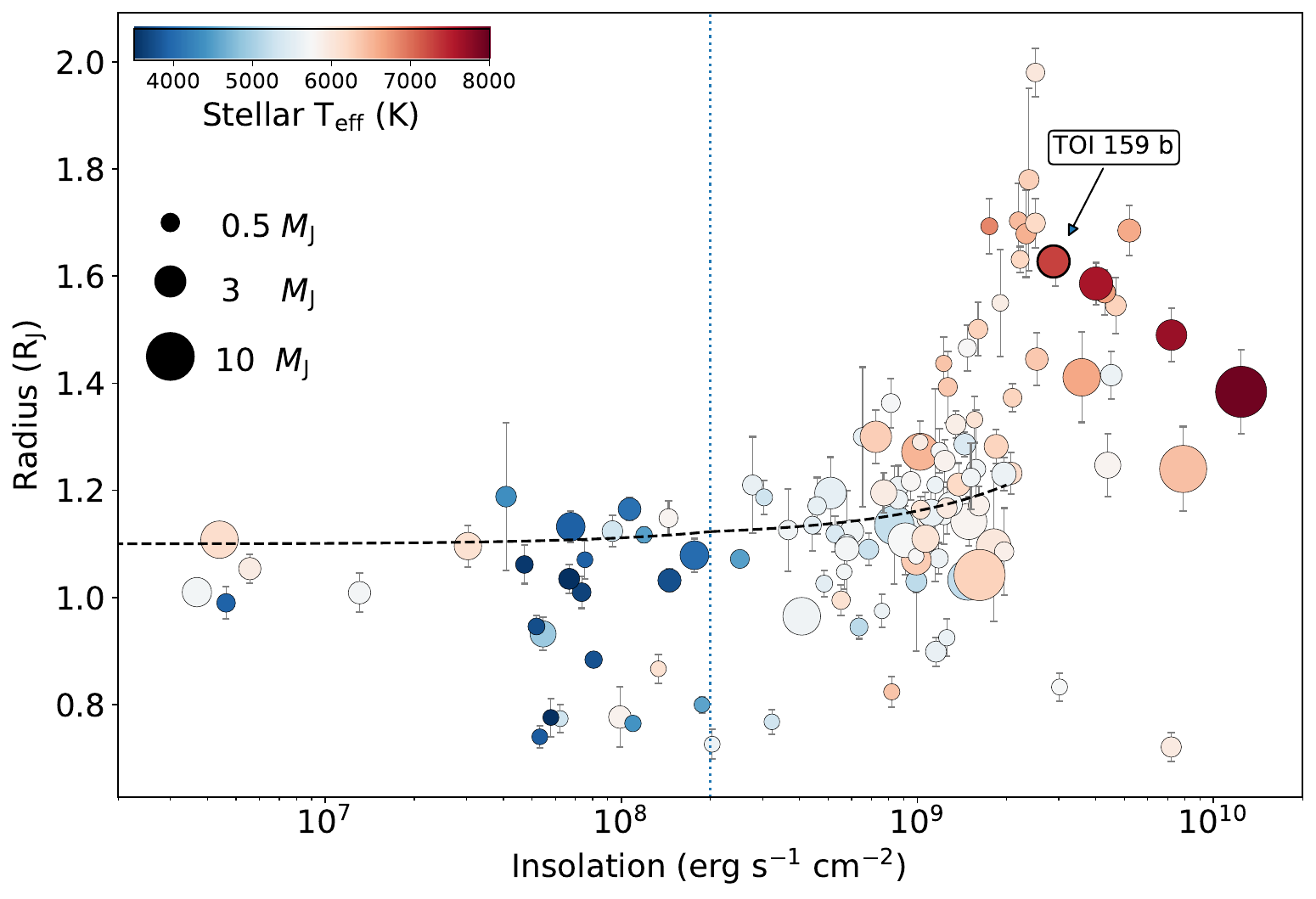}
   \caption{Incident stellar flux versus planet radius. The stellar effective temperature is colour-coded, while the dot size reflects the planetary mass. The black dashed line represents a pure H/He planet (no metals), 4.5 Gyr old, 1 $M_{\rm J}$, without additional internal heating. This figure has been adapted from \cite{2024arXiv240505307T}. We included only planets with masses above 0.2 $M_{\rm J}$. The vertical line corresponds to the inflation threshold at $2 \times 10^8$ erg s$^{-1}$ cm$^{-2}$ \citep{2011ApJS..197...12D, 2016ApJ...831...64T}.}
   \label{fig:infl}
\end{figure}

TOI-159~b is a transiting HJ with a radius of $R_{\rm p} = 1.622 \pm 0.015~R_{\rm J}$ and a mass of $M_{\rm p} = 3.49 \pm 0.26~M_{\rm J}$, which orbits a young $\gamma$ Dor pulsator. Due to its short orbital period and the hot effective temperature of its early F-type host star, TOI-159~b is much more inflated than a pure H/He planet would be without additional extra heating \citep[see Figs. \ref{fig:infl}, \ref{fig:infl_teff} and, for example, ][]{2016ApJ...831...64T}. One possible explanation for such anomalous inflation could be the energy generated by tidal heating because of the planet's eccentric orbit \citep[see next subsection, e.g.][]{2008ApJ...681.1631J}.

\subsection{TOI-159\,b is the hottest eccentric hot-Jupiter}

TOI-159~b joins a rare population of significantly eccentric ($e > 0.1$, $e/\sigma_e > 3$), relatively short-period ($< 5$ d) hot Jupiters, of which WASP-89~b \citep{2015AJ....150...18H}, CoRoT-16~b \citep{2012A&A...541A.149O}, HATS-19~b \citep{2016arXiv160700322B}, HATS-36~b \citep{2018AJ....156..277L}, HAT-P-14~b \citep{2010ApJ...715..458T}, HAT-P-21~b \citep{2011ApJ...742..116B}, HAT-P-31~b \citep{2011AJ....142...95K}, TOI-778~b \citep{2023AJ....165..207C}, and TOI-3593~b \citep{2025ApJS..280...30Y} are the most striking examples. Interestingly, all of the above mentioned planets except TOI-778~b and HAT-P-14~b -- hosted by a star with $T_\textrm{eff} \approx$ 6600 K -- are hosted by solar analogues (5600 $ < T_{\rm eff} < $ 6000 K), while the much higher temperature of TOI-159 ($\approx$ 7300 K) makes it a first. Specifically, it is the hottest HJ ($T_{\rm eq} \simeq 1900$ K) to show a significant orbital eccentricity. However, the sample size is still too small to carry out a population study, and this will be investigated in the future.

Motivated by the significantly eccentric orbit of TOI-159~b, we estimated its circularisation timescale. We did so in four different ways \citep{2006ApJ...649.1004A,2008ApJ...686L..29M, 2008ApJ...678.1396J,2010A&A...516A..64L} and assuming a tidal quality factor $Q'_\textrm{p} = 10^5$, finding a timescale of about $20-70$ Myr, which is shorter than the age of the system. This would suggest that the present eccentricity may require some form of excitation along the lifetime of the system, for example due to a hidden planetary companion. However, by slightly changing the tidal quality factor to $Q'_\textrm{p} = 5\times10^5$, the circularisation becomes longer than the age of the system \footnote{These two alternative findings are also true for the other nine systems with eccentric HJs, where $Q'_\textrm{p} = 10^5$ would not suffice.}. The latter assumption seems reasonable, as short-period HJ tend to have larger $Q'_\textrm{p}$ \citep{2008ApJ...686L..29M, 2017A&A...602A.107B}. 

Following the above findings, which may suggest a possible source of excitation in the eccentricity of TOI-159~b, and encouraged by the observed orbital parameters, we explored whether we could constrain its migration history. The significant eccentricity cannot be explained solely by disc migration \citep{1996Natur.380..606L}; rather it requires a high-eccentricity migration triggered by planet-planet scattering \citep{1996Natur.384..619W, chambers1996, rasioford1996,2006A&A...453..341M} or secular Kozai-Lidov cycles \citep{2003ApJ...589..605W}. We therefore estimated the timescale required for star--planet Kozai--Lidov oscillations \citep{2000ApJ...535..385F} to be triggered by the wide stellar companion TOI-159 B. This would take around $100$ Myr (similar to the age of the system), making this scenario somewhat improbable. Conversely, a planet--planet Kozai--Lidov cycle with a hypothetical giant planet at 10 au would result in a timescale that is an order of magnitude shorter. Such a scenario would require the presence of a second planet in a highly eccentric and relatively nearby orbit, which however is not currently detectable with the limited number of RVs available.

\subsection{Well-characterised planets orbiting early-type stars}
According to the PlanetS catalogue \citep{2020A&A...634A..43O, 2024A&A...688A..59P}, a total of 143 giant ($\ge$ 7 $R_\mathrm{\oplus}$) planets with well-constrained densities ($\sigma_\rho/\rho < 20\%$) and transiting early-type stars above the Kraft break \citep[i.e. stars with $T_\textrm{eff} \gtrsim 6200~\textrm{K}$,][]{1967ApJ...150..551K} have been discovered as of December 2025. It is worth emphasising that the TESS mission largely contributed to expand this sample, with more than 50 discoveries to date. TOI-159~b joins this limited group of planets with well-characterised properties orbiting hot stars, and becomes the twelfth planet whose host star has $T_\textrm{eff} > 7000~\textrm{K}$, and the fourth such system discovered by TESS.

\subsection{S-type planets}
An ``S-type'' planet, such as TOI-159~b, orbits one star in a binary system \citep[e.g.][]{2025A&A...696A..86S}. In such systems, the gravitational influence of the companion can truncate the protoplanetary disc outward, limiting the material for planet formation \citep{2021MNRAS.504.2235Z}. This effect is stronger in close binaries \citep[i.e. those with separation $<$ 1000 au][]{1994ApJ...421..651A}, making them natural laboratories for planet formation models \citep{2021FrASS...8...16F, 2023AJ....165..177S}. We expect these systems to show the largest differences in planet properties compared to single stars \citep{2021AJ....161..134H, 2016AJ....152....8K}, as the stellar companion can influence both the formation and evolution of planets \citep[e.g.][]{2015pes..book..309T, 2021MNRAS.507.3593M}. For example, the reduced mass and lifetime of the disc may shorten the time available for a planetary core to accrete and maintain a gaseous atmosphere \citep[e.g.][]{2012ApJ...745...19K}, reduce the mass budget of solids, or alter the distribution of solids \citep[e.g.][]{2023ASPC..534..717D, 2024AJ....168..129S}.

TOI-159 b~is only the sixth S-type planet in a close-binary system orbiting a hot ($T_{\rm eff} > 7000$K) star, compared to $>50$ reported discoveries for single-star planets \citep[see NASA Exoplanet Archive,][]{2025arXiv250603299C}. 

\subsection{Atmospheric retrieval} \label{sec:atmoretrieval}

To date the atmospheric characterisation of hot Jupiters transiting pulsating stars within the literature is sparse \citep[e.g.][]{haynes2015, zhang2018, changeat2022} likely resulting from both the limited number of known systems and additional challenges posed by the host star. Equally, these planets have been largely analyzed in emission or through phase curves, so the IMACS TOI-159 b transmission spectrum has the potential to offer a different, unique perspective on their atmospheres. 

We attempted to extract information about the planetary atmosphere from this spectrum through a small-scale, preliminary retrieval experiment. Multiple potential models were explored following a hierarchical approach beginning with that of a grey model i.e. consistent with a flat spectrum and iteratively adding model contributions in the form of active chemistry and/or stellar contamination. We use the fully Bayesian, open-source retrieval code \texttt{TauREx\,3} \citep{2021ApJ...917...37A,2022ApJ...932..123A} enhanced with the stellar activity plugin \texttt{ASteRA} \citep{thompson2024}. Combined star-planet retrievals are leveraged as the $\log R'_{HK}$ index we recover is indicative of moderate chromospheric activity. Additionally, early studies of $\gamma$ Dor pulsators showed that photometric observations could be reproduced through the interaction of starspots and differential rotation \citep{balona1994}, although this explanation was subsequently ruled out. These reasons motivate our exploration of the host star potentially contributing to the observed spectrum via starspot-driven contamination.

\begin{figure}
   \centering
   \includegraphics[width=\hsize]{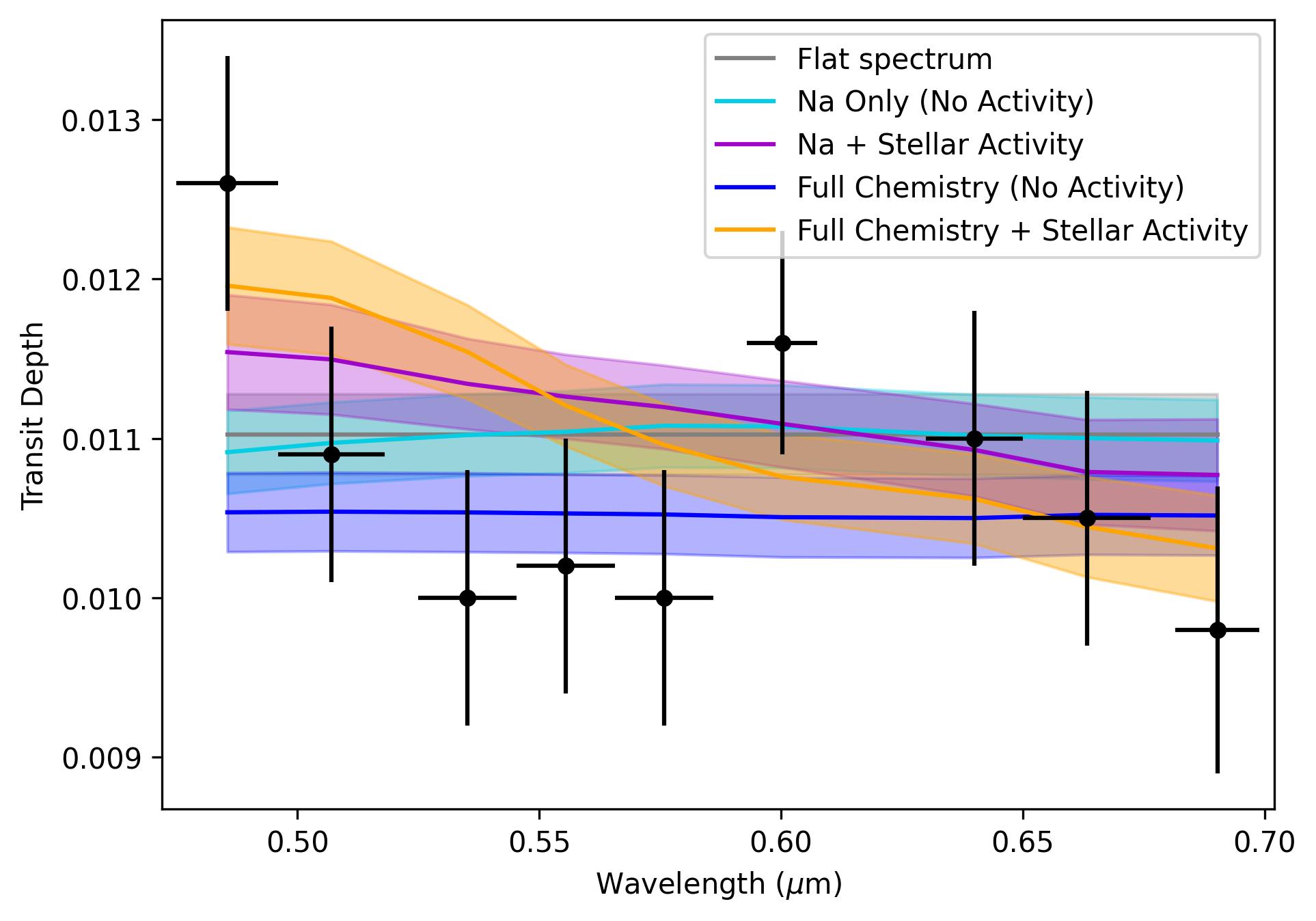}
   \caption{Best fit retrieved spectra for 4 of the retrieval models explored. The models are visually distinct from one another despite having equivalent Bayesian evidences.}
   \label{fig:bestfit_spec}
\end{figure}

The best fit spectra for five of the retrieval models explored are shown in Fig.~\ref{fig:bestfit_spec}. More detailed information about the retrievals conducted and their results can be found in Appendix~\ref{sec:retrieval+}. Unfortunately, the resolution of our observation is limiting and the models explored are statistically indistinguishable from one another in terms of Bayesian evidence, meaning that no model is definitively preferred at this stage and no conclusive feature detection can be claimed. These degenerate models highlight the need for follow up observations at higher resolution and/or with a wider wavelength coverage to confidently characterise the atmosphere of TOI-159\,b for example with the James Webb Space Telescope (JWST) (Appendix~\ref{sec:jwst_prospects}). If stellar contamination is indeed contributing to the observed opacity, then the planetary radius may be significantly smaller at $\sim$1.45 $R_{\rm Jup}$. A smaller radius would have implications for mass inflation models and bring TOI-159\,b closer to that predicted by existing mass-radius relations. However, we stress that contamination from pulsating stars and from photospheric heterogeneities is not guaranteed to be equivalent. A pulsation-specific contamination model would be required to robustly explore all potential scenarios.

\subsection{Rossiter–McLaughlin measurement prospects}

The study of the Rossiter-McLaughlin (RM) effect \citep[e.g.][]{2000A&A...359L..13Q}, which has not yet been explored for this system, could constrain the true obliquity ($\psi$) angle and shed light on the orbital architecture and migration mechanism of TOI-159~b \citep[e.g.][]{2011Natur.473..187N}. Such studies are vital for improving our understanding of planets transiting massive, early-type stars \citep{2023A&A...675A..39P}. The RV anomaly caused by the RM effect is expected to be $\simeq 230$ m s$^{-1}$ \citep{2010exop.book...55W}, which is about ten times larger than the RV uncertainties of our HARPS data. Observations of young transiting planets, which have not experienced significant tidal alterations, provide a unique opportunity to study their initial obliquity configuration \citep[e.g.][]{2022PASP..134h2001A, 2024A&A...684L..17M}. We estimated the radiative tidal realignment timescale following \cite{2012ApJ...757...18A}, and found a decay timescale for this system exceeding the Hubble time. This implies that the obliquity angle should provide a direct diagnostic of its primordial formation pathway.

%%%%%%%%%%%%%%%%%%%%%%%%%%%%%%%%%%%%%%%%%%%%%%%%%%%%%%%%%%%%%%
\section{Conclusions}
\label{sec:conclusions}

We presented the discovery of the eccentric giant planet TOI-159~b orbiting a young, pulsating $\gamma$ Doradus star. Using HARPS and CORALIE RVs jointly modelled with \textit{TESS} and IMACS photometry, we were able to measure its mass, disentangle the stellar rotational modulation and pulsation periods, and generate a low-resolution transmission spectrum.

We emphasise the key findings that emerged from this study:
\begin{itemize}
    \item TOI-159 is young ($\approx$ 150 Myr) and active early F-type star, which appears to be a classical $\gamma$ Dor pulsator with two main non-radial g-modes; 
    \item The giant planet TOI-159~b is anomalously inflated, possibly driven by tidal heating due to its short-period, eccentric orbit;
    \item TOI-159~b is the hottest ($T_{\rm eq} \simeq 1900$ K) currently known HJ to show a significant eccentricity. This suggests a migration history involving post-formation dynamical interactions like planet-planet scattering or secular Kozai-Lidov cycles;
    \item It is only the sixth S-type planet discovered orbiting a hot star ($T_\textrm{eff} > 7000~\textrm{K}$). This system may hence serve as a laboratory for planet formations studies, as the companion may have influenced the protoplanetary disc evolution, limiting the mass available for planet accretion;
    \item The IMACS transmission spectrum appears to show potential features in the form of a positive blueward slope and a broad absorption feature. Although we explored the possibility of starspot-driven contamination, the limiting resolution resulted in degenerate solutions. This highlights the need for high-resolution follow-up studies to confidently characterise TOI-159~b's atmosphere. If the observed bluewards slope is indeed caused by stellar contamination, the true planetary radius could be less inflated.  
\end{itemize}

%%%%%%%%%%%%%%%%%%%%%%%%%%%%%%%%%%%%%%%%%%%%%%%%%%%%%%%%%%%%%%
\begin{acknowledgements}
We acknowledge the use of public TESS data from pipelines at the TESS Science Office and at the TESS Science Processing Operations Center. Funding for the TESS mission is provided by NASA's Science Mission Directorate. KAC acknowledges support from the TESS mission via subaward s3449 from MIT. This research has made use of the Exoplanet Follow-up Observation Program (ExoFOP; DOI: 10.26134/ExoFOP5)  website, which is operated by the California Institute of Technology, under contract with the National Aeronautics and Space Administration under the Exoplanet Exploration Program. S.V. gratefully acknowledges the support provided by Fondecyt Regular n. 1220264 and by the ANID BASAL project FB210003. This publication has been made possible by Spanish grants PID2021-125627OB-C31 and PID2024-158486OB-C31 funded by MCIU/AEI/10.13039/501100011033 and by “ERDF A way of making Europe”, by the programme Unidad de Excelencia María de Maeztu CEX2020-001058-M financed by MCIN/AEI/10.13039/501100011033 and by the MaX-CSIC Excellence Award MaX4-SOMMA-ICE, by the Generalitat de Catalunya/CERCA programme, and by the European Research Council (ERC) under the European Union’s Horizon Europe programme (ERC Advanced Grant SPOTLESS; no. 101140786). Views and opinions expressed are however those of the author(s) only and do not necessarily reflect those of the European Union or the European Research Council. Neither the European Union nor the granting authority can be held responsible for them. This work was carried out within the framework of the Swiss National Centre for Competence in Research (NCCR) PlanetS supported by the Swiss National Science Foundation (SNSF) under grants 51NF40$\_$182901 and 51NF40$\_$205606. This work makes use of observations from the LCOGT network. Part of the LCOGT telescope time was granted by NOIRLab through the Mid-Scale Innovations Program (MSIP). MSIP is funded by NSF. V.~N.~and G.~P.~acknowledge financial support from the Bando Ricerca Fondamentale INAF 2023, Data Analysis Grant: ``Characterization of transiting exoplanets by exploiting the unique synergy between TASTE and TESS''. G.~P.~ and G.~M.~acknowledge support by the Space It Up project funded by the Italian Space Agency, ASI, and the Ministry of University and Research, MUR, under contract n. 2024-5-E.0 - CUP n. I53D24000060005. ML acknowledges support of the Swiss National Science Foundation under grant number PCEFP2\_194576. AT acknowledges support from ESA through the Science Faculty - Funding reference 4000146542.

\end{acknowledgements}

% Please note that we have included the references below in
% order to compile the document, but we ask you to:
\bibliographystyle{aa} % style aa.bst
\bibliography{references} % your references 

%%%%%%%%%%%%%%%%%%%%%%%%%%%%%%%%%%%%%%%%%%%%%%%%%%%%%%%%%%%%%%%
% Appendices must be placed after   \end{thebibliography}
% They will be placed automatically on a new page.
%%%%%%%%%%%%%%%%%%%%%%%%%%%%%%%%%%%%%%%%%%%%%%%%%%%%%%%%%%%%%%%
\begin{appendix}
%%%%%%%%%%%%%%%%%%%%%%%%%%%%%%%%%%%%%%%%%%%%%%%%%%%%%%%%%%%%%%%
\section{Independent stellar parameters determination}
As an independent determination of the basic stellar parameters, we performed an analysis of the broadband SED of the star together with the {\it Gaia\/} DR3 parallax \citep[with no systematic offset applied; see, e.g.,][]{StassunTorres:2021}, to determine an empirical measurement of the stellar radius, following the procedures described in \citet{Stassun:2016,2017AJ....153..136S,Stassun:2018}. We pulled the $JHK_S$ magnitudes from {\it 2MASS}, the $G_{\rm BP}$, $ G_{\rm RP}$ magnitudes from {\it Gaia}, and the W1--W4 magnitudes from {\it WISE}. We also utilized the absolute flux-calibrated spectrophotometry from {\it Gaia}. Together, the available photometry spans the full stellar SED over the wavelength range 0.4--20~$\mu$m (see Figure~\ref{fig:sed}).  

We performed a fit using PHOENIX stellar atmosphere models \citep{2013A&A...553A...6H}, with $T_{\rm eff}$, $\log g$, and [Fe/H] adopted from the spectroscopic analysis. The extinction, $A_V$, was limited to maximum line-of-sight value from the Galactic dust maps of \citet{Schlegel:1998}. The resulting fit (Figure~\ref{fig:sed}) has a reduced $\chi^2$ of 2.0, with a best-fit $A_V = 0.50 \pm 0.03$. Integrating the (unreddened) model SED gives the bolometric flux at Earth, $F_{\rm bol} = 1.658 \pm 0.039 \times 10^{-9}$ erg~s$^{-1}$~cm$^{-2}$. Taking the $F_{\rm bol}$ together with the {\it Gaia\/} parallax directly gives the bolometric luminosity, $L_{\rm bol} = 6.28 \pm 0.15$~L$_\odot$. The stellar radius follows from the Stefan-Boltzmann relation, giving $R_\star = 1.571 \pm 0.036$~R$_\odot$. In addition, we can estimate the stellar mass from the empirical relations of \citet{Torres:2010}, giving $M_\star = 1.64 \pm 0.10$~M$_\odot$. Finally, we can estimate the projected rotation period from the spectroscopic $v\sin i$ and $R_\star$, giving $P_{\rm rot}/\sin i =  3.61 \pm 0.34$~d. It is important to note that these estimated parameters agree well with those presented in Sect. \ref{sec:irfm} and the measured values (see Table \ref{tab:star_param}).

\section{Search for transit timing variations}
To determine whether dynamical interactions are present in the system, we searched for transit timing variations \citep[TTVs; ][]{2005MNRAS.359..567A, 2005Sci...307.1288H, 2014A&A...571A..38B} of TOI-159~b using a \texttt{PyORBIT} model based on \texttt{BATMAN}. This specific model allows us to fit each transit time ($T_0$) independently, while keeping the orbital period fixed at the value found in Sect. \ref{sec:analysis}. The optimisation algorithms and the convergence criteria are the same applied in the previous (non-TTV) analysis and described in Section~\ref{sec:modelling}.

We produced the observed$-$calculated $(O-C)$ diagram for TOI-159~b (see Fig. \ref{fig:ttv}) after removing our best-fit transit ephemeris from our individually fitted transit times. Even though there seems to be a slight offset between different sectors, the overall reduced $\chi^2$ from a straight line fit is very close to one and a GLS periodogram calculated on the $(O-C)$ values yields no power peak at high significance. We conclude that no coherent TTV signal can be detected down to an amplitude consistent with our scatter ($\simeq 1$~min), and that small offsets can be easily justified with a combination of stellar activity and residuals from the fit of stellar pulsations.

\section{Ephemeris improvements and future observations}

Propagating the new ephemeris to 1 January 2030 significantly reduces the level of uncertainty for TOI-159~b to 0.3 minutes. This corresponds to a 92\% decrease from the previous value of 4 minutes, which was obtained by propagating the ephemeris available on the ExoFOP website. 

\section{Extra figures and tables}
\begin{figure}[h]
   \centering
   \includegraphics[width=\hsize]{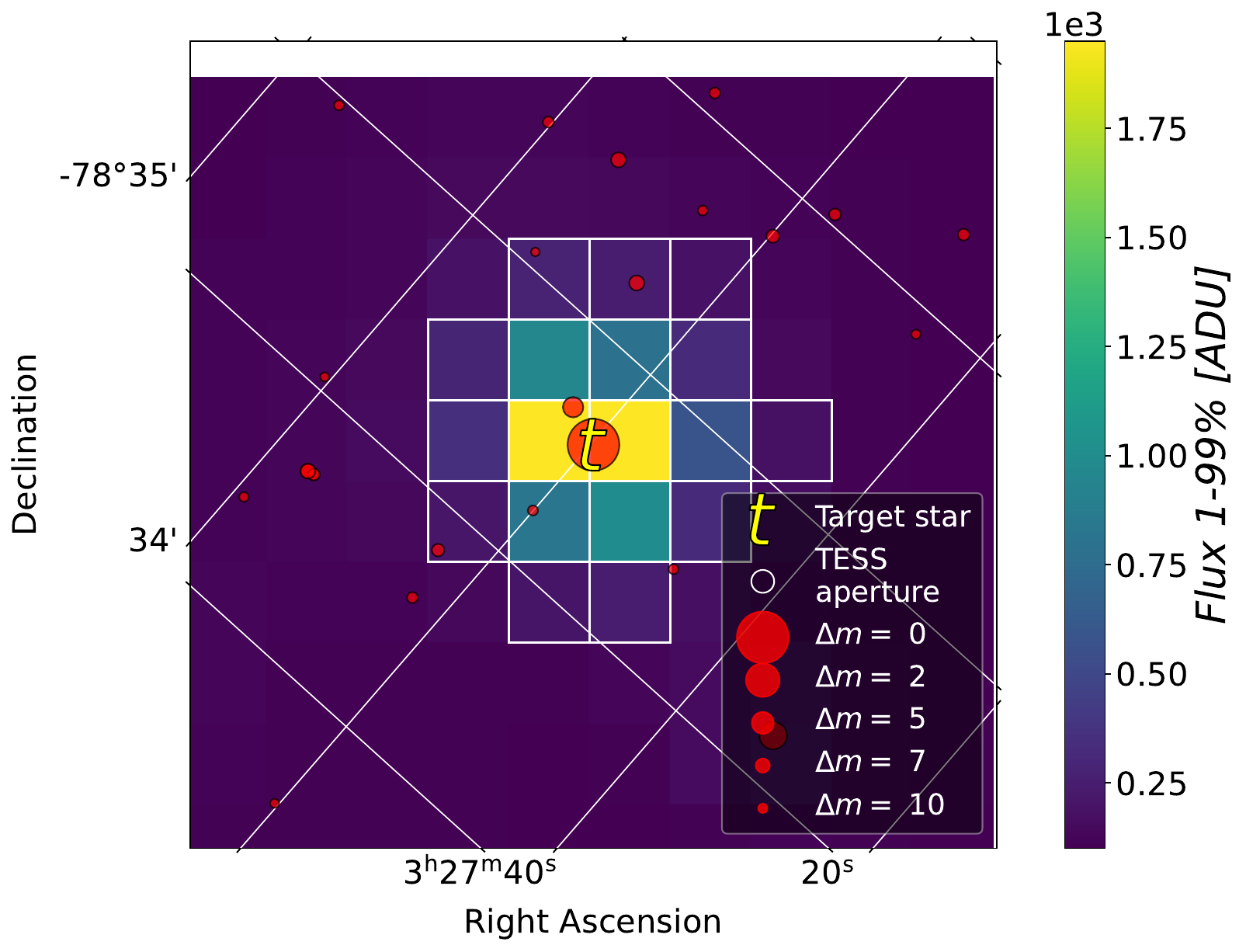}
   \caption{\textit{Gaia} stars identified in the \textit{TESS} field. The yellow letter `t' highlights the position of TOI-159. None of the \textit{Gaia} stars can be a blended eclipsing binary. }
   \label{fig:field}
\end{figure}

\begin{figure}[h]
   \centering
   \includegraphics[trim={3.5cm 2.5cm 2cm 2.5cm},width=\hsize]%
   {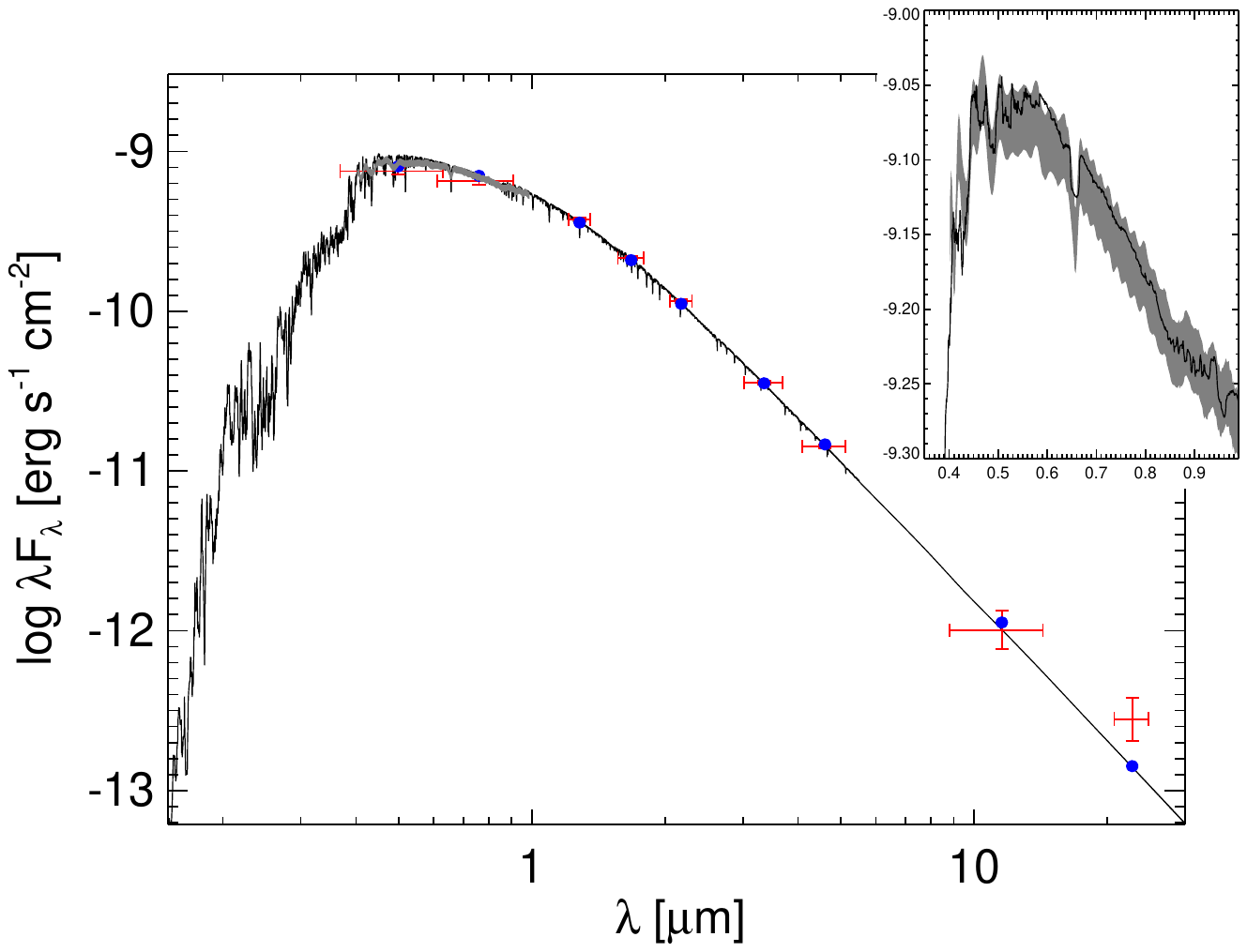}
   \caption{Spectral energy distribution of TOI-159. Red symbols represent the observed photometric measurements, where the horizontal bars represent the effective width of the passband. Blue symbols are the model fluxes from the best-fit PHOENIX atmosphere model (black). The inset shows the {\it Gaia\/} spectrophotometry as a gray swathe overlaid on the best-fit model (black).}
   \label{fig:sed}
\end{figure}

\begin{figure}[h]
   \centering
   \includegraphics[width=\hsize]%
   {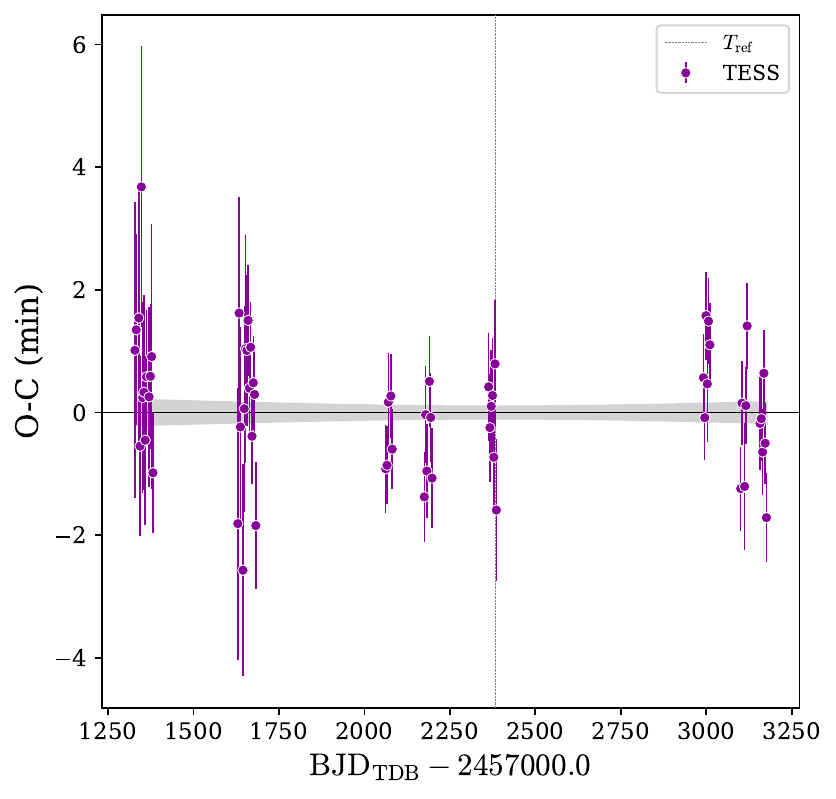}
   \caption{O$-$C plot displaying the observed (O) minus calculated (C) transit times, relative to the linear ephemeris of TOI-159~b.}
   \label{fig:ttv}
\end{figure}

\begin{figure}
   \centering
   \includegraphics[width=\hsize]%
   {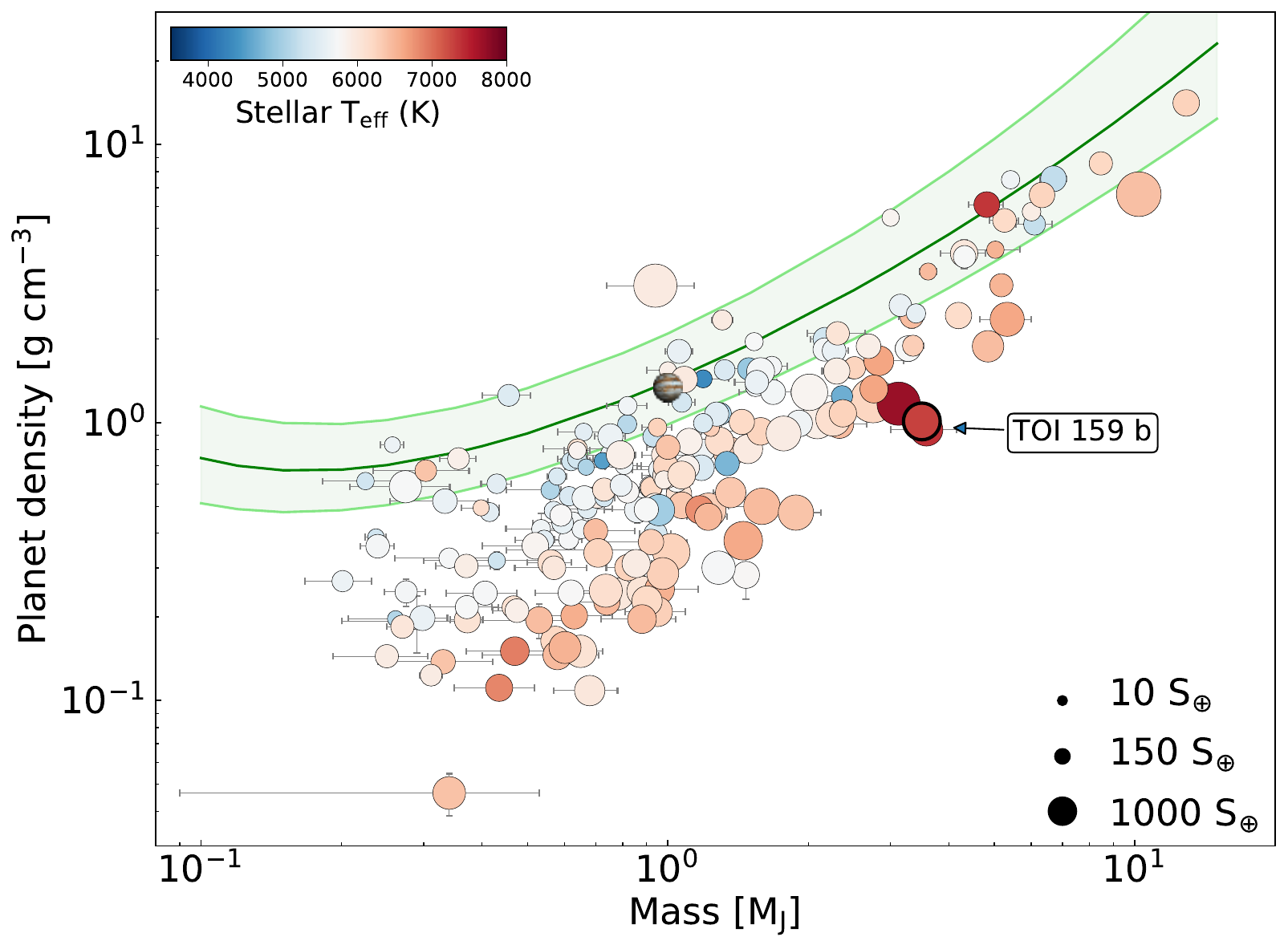}
   \caption{Planet bulk density as a function of planetary mass of well-characterised irradiated ($F_\star > 2 \times 10^8$ erg s$^{-1}$ cm$^{-2}$) giant planets. The stellar $T_\textrm{eff}$ is colour-coded, while the size of the dots reflects the planet insolation. The green shaded area highlights the mass--radius relation for giant planets in the absence of anomalous heating, that is when assuming no inflation \citep{2024arXiv240505307T}. Adapted from \cite{2024A&A...691A..67M}.}
   \label{fig:infl_teff}
\end{figure}

\begin{figure*}[h]
   \centering
   \sidecaption
   \includegraphics[width=12cm]{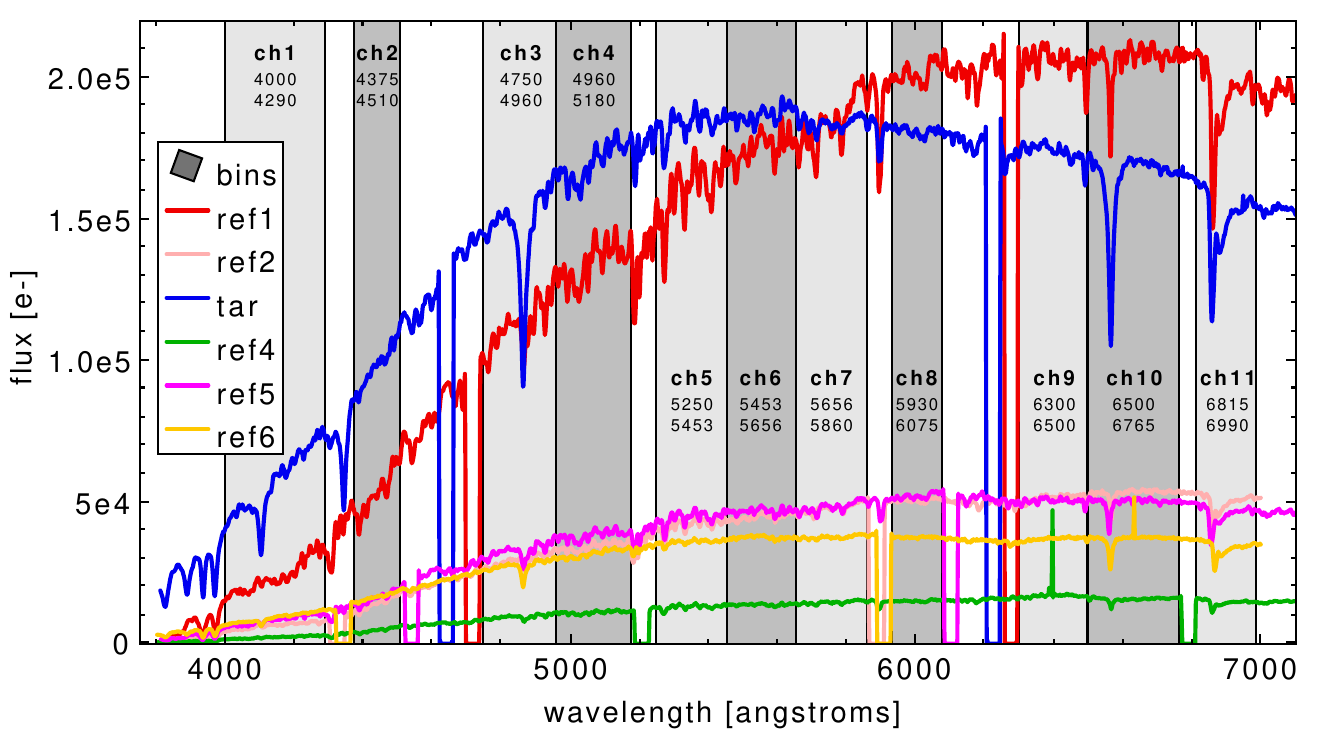}
   \caption{Sample spectra of TOI-159 (blue line) and of the five reference stars gathered with IMACS (color-coded according to the legend), see details in Sec.~\ref{sec:imacs}. The vertical gray bands mark  the spectro-photometric bins applied to the IMACS data to extract the transmission spectrum of TOI-159.  Each band is labelled with the bin identification number and its wavelength range in micron. }
   \label{fig:channels}
\end{figure*}

\begin{figure}
   \centering
   \includegraphics[width=\hsize]{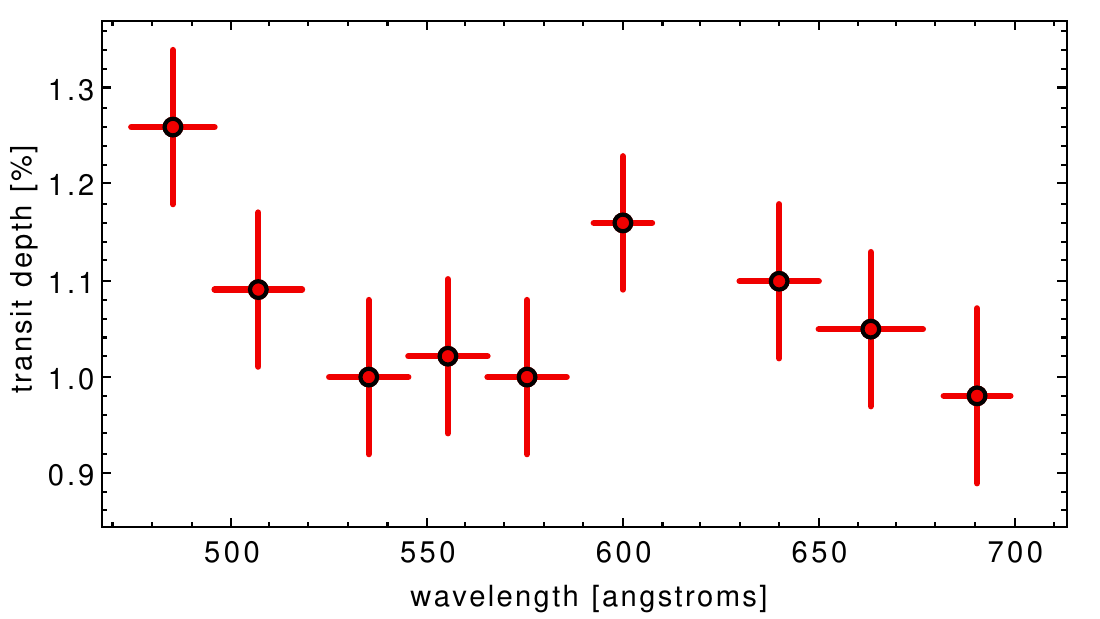}
   \caption{Transmission spectrum of TOI-159 derived from our IMACS data, plotted as transit depth $(R_{\rm p}/R_\star)^2$ as a function of wavelength. The horizontal error bars correspond to the spectral range of the spectrophotometric channels extracted (\#3 to \#11 in Fig.~\ref{fig:channels}; see Section~\ref{sec:imacs} for details).}
   \label{fig:transmission_spectrum}
\end{figure}

\begin{figure}
    \centering
    \includegraphics[width=0.45\textwidth]{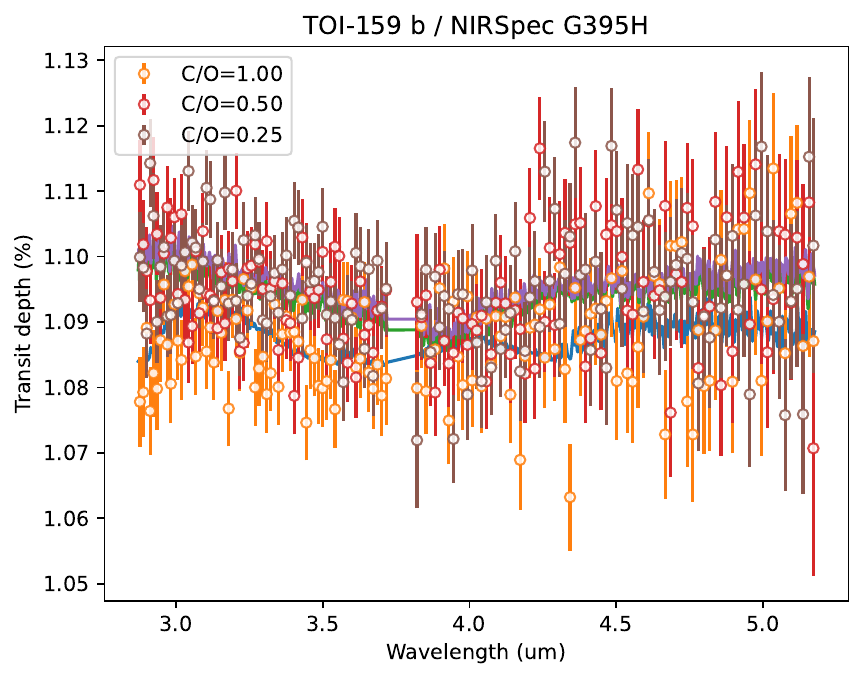}
    \caption{Simulated JWST/NIRSpec BOTS G395H+F290LP transmission spectra for TOI-159b assuming equilibrium chemistry at solar metallicity and three C/O ratios (0.25, 0.5, 1.0). The points and error bars show the expected measurement precision from the time-series simulator, while the solid curves indicate the underlying forward models. Within the predicted uncertainties, the three C/O scenarios are largely degenerate over the G395H bandpass.}
    \label{fig:jwst_g395h_three_co}
\end{figure}

\begin{table}[]
    \centering
    \caption{Tabular format of TOI-159~b transmission spectrum. }
    \begin{tabular}{ccccc}\hline\hline
ch.\tablefootmark{a} & $\lambda_1$~[\AA{}] & $\lambda_2$~[\AA{}] & $(R_{\rm p}/R_\star)^2$ & $\sigma(R_{\rm p}/R_\star)^2$ \\ \hline
3  &4750&  4960&  0.0126&  0.0008\\
4  &4960&  5180&  0.0109&  0.0008\\
5  &5250&  5453&  0.0100&  0.0008\\
6  &5453&  5656&  0.0102&  0.0008\\
7  &5656&  5860&  0.0100&  0.0008\\
8  &5930&  6075&  0.0116&  0.0007\\
9  &6300&  6500&  0.0110&  0.0008\\
10 &6500&  6765&  0.0105&  0.0008\\
11 &6815&  6990&  0.0098&  0.0009\\ \hline
    \end{tabular}
    \tablefoot{\tablefoottext{a}{Channel ID number.}}
    \label{tab:trans}
\end{table}

\section{Retrieval experiment: Additional information}\label{sec:retrieval+}

\begin{figure}
   \centering
   \includegraphics[width=\hsize]{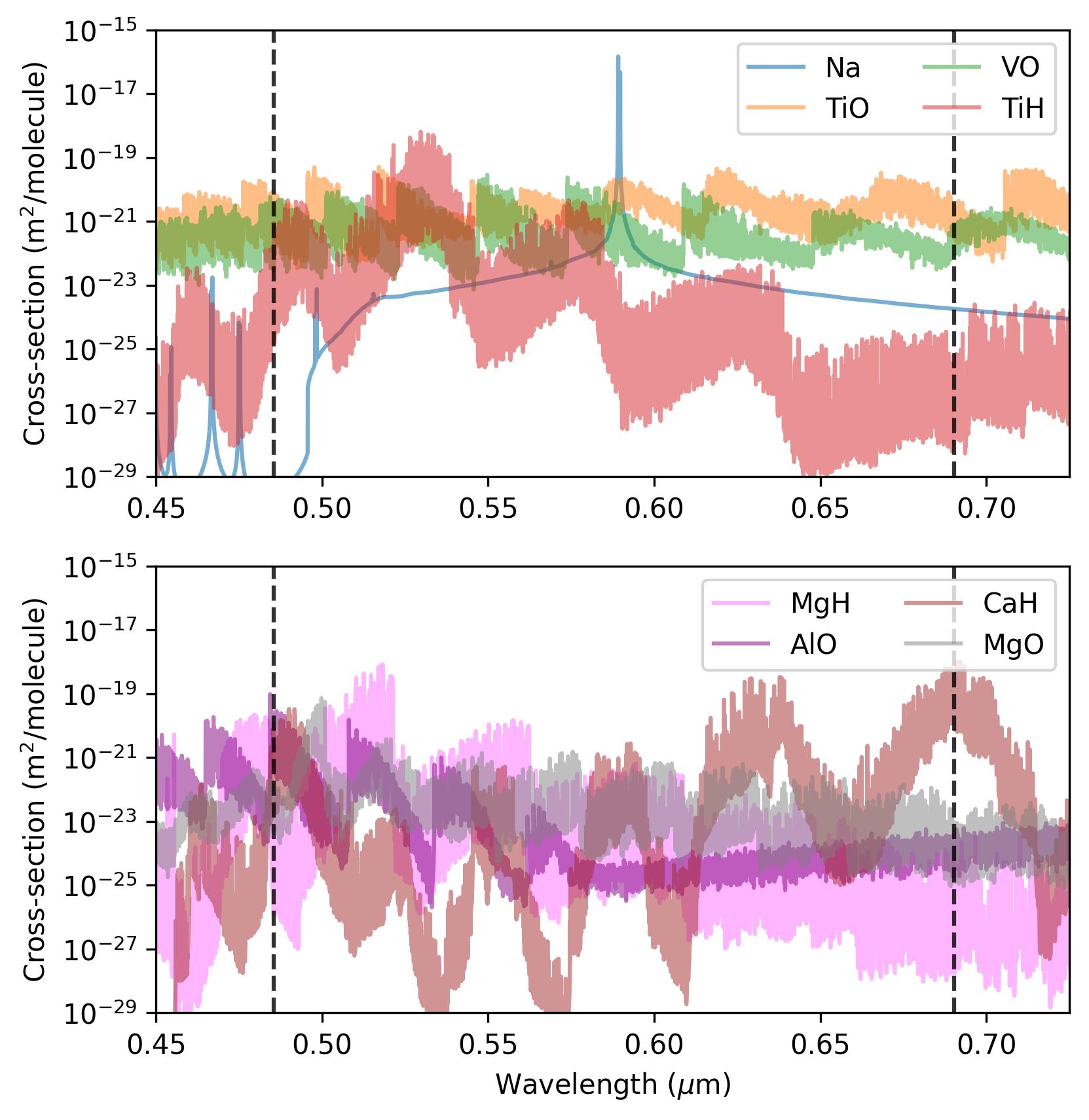}
   \caption{ExoMol cross sections for the molecules (and atom) considered in our retrieval study at 1900 K consistent with the derived equilibrium temperature of TOI-159\,b. The black dashed lines indicate the lower and upper wavelength bounds of IMACS respectively. The molecules have been separated across two plots for clarity but the axes scales remain the same for both.}
   \label{fig:molecular_opacities}
\end{figure}

We consulted the ExoMol database \citep{tennyson2016,tennyson2024} to identify several atomic and molecular species with significant opacity in the optical regime probed by IMACS. We deemed eight species as having substantial opacity to warrant exploration: Na \citep[e.g.][]{allard2019}, TiO \citep[e.g.][]{mckemmish2019}, VO \citep[e.g.][]{mckemmish2024}, TiH \citep{burrows2005}, MgO \citep[e.g.][]{li2019}, MgH \citep{owens2022}, CaH \citep{owens2022} and AlO \citep[e.g.][]{patrascu2015}. The cross sections for these molecules at the relevant wavelengths are given in Fig.~\ref{fig:molecular_opacities}. Although widely explored within the literature as a potential optical absorber, we neglected to include K within our retrieval models, as its most prominent doublet feature lies outside the IMACS wavelength coverage at 0.77 $\mu$m. Absorption from Rayleigh scattering and collision-induced absorption (H$_2$-H$_2$ and H$_2$-He pairs) are always considered.

Stellar contamination can also be a substantial source of optical opacity if present. The presence of active regions on the stellar surface is well known to introduce highly chromatic contamination effects in transmission spectra by breaking the homogeneity of the host star as the light source. Active regions are characterised by distinct spectral energy distributions resulting in wavelength-dependent contamination which strengthens exponentially as a function of decreasing wavelength imparting a characteristic positive bluewards slope to the optical spectrum. Such a slope can obscure features of other optical absorbers and be easily confused with Rayleigh scattering in the planetary atmosphere \citep[e.g.][]{ballerini2012, thompson2024}. 
Scenarios accounting for potential stellar contamination were modelled with \texttt{ASteRA} \citep{thompson2024} which supports combined star-planet retrievals within \texttt{TauREx\,3} through the inclusion of a simplified stellar model; photospheric heterogeneities e.g. spots and faculae are parameterised by their respective temperatures and filling factors respectively. In this scenario we only consider spots, motivated by the potential slope visible in the transmission spectrum and the desire to minimise the dimensionality of the retrieval models to protect against overfitting.

The range of fitting parameters considered in this retrieval experiment and their respective priors are given in Table~\ref{tab:retr_fitparams}. Wide, unrestrictive priors are chosen to prevent biasing the retrieval optimiser. All retrievals use the nested sampling algorithm MultiNest \citep{feroz2009} as the optimiser with 1000 live points and an evidence tolerance of 0.5.
The highest dimensionality retrieval conducted explored both the possibility of contamination via stellar activity and all 8 molecules of interest in a free chemistry retrieval which we term ‘Full Chemistry’. With 12 fitting parameters, this model is overly complex given the number of IMACS data points but was conducted to identify which, if any, of these molecules are more likely to be present. A variety of lower dimensionality retrievals were also conducted following a hierarchical framework, sequentially varying the combination of molecules and physical processes in efforts to determine their relative importances beginning from a grey model assuming a completely flat spectrum i.e. no transmission features. The results are given in Table~\ref{tab:retr_results} although it soon became apparent that the resolution is unfortunately not high enough to confidently support or reject any of the models considered.

\begin{table}
\centering
\caption{Fitting parameters and their priors for the retrievals conducted in Section~\ref{sec:atmoretrieval}. }
\renewcommand{\arraystretch}{1.5}
\begin{tabular}{lcc}
\hline
Parameter & Prior Bounds & Scale \\ \hline \hline
$R_{\rm p}$ ($R_{\rm Jup}$)   & 0.5 ; 3  & linear \\            
$T_{\rm p}$ (K)   & 500 ; 3500    & linear \\             
log($X$)\tablefootmark{a}   & -12 ; -1  & log$_{10}$ \\ \hline
$T_{\rm spot}$\tablefootmark{b}  & 3000 ; 7250   & linear \\
$F_{\rm spot}$\tablefootmark{b}  & 0 ; 0.99  & linear \\ \hline
\end{tabular}
\label{tab:retr_fitparams} 
\tablefoot{\tablefoottext{a}{Volume mixing ratio of all chemical species considered.} \tablefoottext{b}{Spot parameters which are only fit for with \texttt{ASteRA}.}}
\end{table}

\begin{table*}
\centering
\caption{Complete retrieval results for 6 retrieval models of varying complexity: a grey, flat-line model, 1) Na only; 2) Stellar activity only; 3) Na + stellar activity; 4) Full chemistry (no stellar activity); 5) Full chemistry + stellar activity. }
\renewcommand{\arraystretch}{1.2}
\resizebox{\linewidth}{!}{%
\begin{tabular}{lccccccccccccc}
\hline
Model & $R_{\rm p}$ ($R_{\rm Jup}$) & $T_{\rm p}$ (K) & log(Na) & log(TiO) & log(VO) & log(TiH) & log(MgH) & log(MgO) & log(AlO) & log(CaH) & $T_{\rm spot}$ (K) & $F_{\rm spot}$ & ln(z) \\ \hline \hline
grey & 1.61 $\pm^{0.02}_{0.03}$ & 1920 $\pm^{1100}_{940}$ & - & - & - & - & - & - & - & - & - & - & 46.7 \\ 
1 & 1.62 $\pm$ 0.02 & 1970 $\pm^{1040}_{1000}$ & -6.8 $\pm^{3.7}_{3.5}$ & - & - & - & - & - & - & - & - & - & 46.8 \\ 
2 & 1.50 $\pm^{0.09}_{0.16}$ & 1900 $\pm^{1010}_{970}$ & - & - & - & - & - & - & - & - & 5540 $\pm^{860}_{1590}$ & 0.43 $\pm^{0.32}_{0.26}$ & 47.2 \\
3 & 1.51 $\pm^{0.08}_{0.20}$ & 2010 $\pm^{960}_{990}$ & -7.5 $\pm^{3.9}_{3.0}$ & - & - & - & - & - & - & - & 5700 $\pm^{950}_{1910}$ & 0.39 $\pm^{0.35}_{0.26}$ & 47.2 \\
4 & 1.60 $\pm$ 0.03 & 1940 $\pm^{900}_{870}$ & -6.6 $\pm^{3.5}_{3.2}$ & -6.9 $\pm^{3.1}_{2.9}$ & -6.8 $\pm^{3.6}_{3.1}$ & -6.6 $\pm^{3.2}_{3.4}$ & -6.3 $\pm^{3.3}_{3.4}$ & -6.3 $\pm 3.4$ & -6.4 $\pm^{3.1}_{3.2}$ & -6.8 $\pm3.3$ & - & - & 47.0 \\
5 & 1.44 $\pm^{0.10}_{0.13}$ & 2030 $\pm880$ & -6.7 $\pm 3.1$ & -6.5 $\pm^{3.2}_{3.1}$ & -6.7 $\pm^{3.1}_{3.0}$ & -7.2 $\pm^{3.4}_{2.8}$ & -6.5 $\pm 3.2$ & -6.3 $\pm^{3.1}_{3.2}$ & -6.6 $\pm^{3.2}_{3.3}$ & -7.4 $\pm^{3.6}_{2.9}$ & 5370 $\pm^{840}_{1280}$ & 0.43 $\pm 0.26$ & 47.1 \\ \hline
\end{tabular}%
}
\label{tab:retr_results} 
\end{table*}

\begin{figure*}
   \centering
   \includegraphics[width=\hsize]{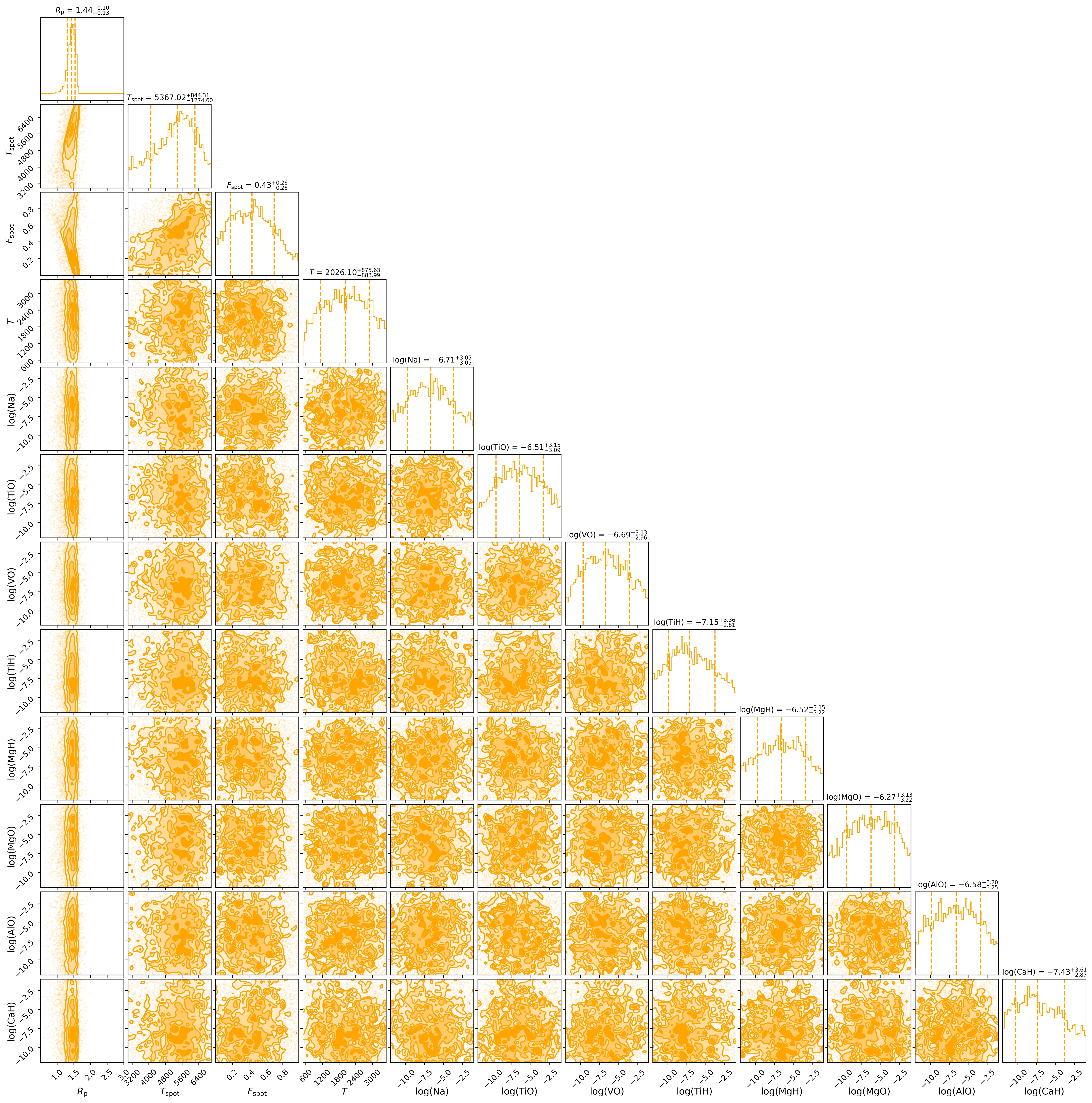}
   \caption{The posterior distribution retrieved for Model 5: Full chemistry and stellar activity.}
   \label{fig:retr_posterior}
\end{figure*}

The best fit \texttt{ASteRA} models are visually distinct from the other retrieval models tested, preferring an underlying positive bluewards slope (Fig.~\ref{fig:bestfit_spec}),
although we again stress that the slight increase in the Bayesian evidence recovered for contaminated models is too low to be deemed statistically significant by interpretation scales such as the Jeffrey's scale \citep[e.g.][]{trotta2008}. Some constraints can be placed on the best-fit spot parameters with the posterior distribution (Fig.~\ref{fig:retr_posterior}) indicating a slight preference for spots with temperatures $\sim$ 2000 K cooler than the quiet photosphere covering around 40\% of the visible stellar hemisphere. No model is able to reproduce the modulation at $\sim$ 0.6 $\mu$m and no chemical species can be confidently detected at this resolution, as evidenced by the posterior distribution (Fig.~\ref{fig:retr_posterior}).

\section{Prospects for JWST atmospheric follow-up}
\label{sec:jwst_prospects}

This system offers an interesting opportunity for atmospheric characterization with JWST. TOI-159b is an inflated, massive hot Jupiter on a short-period but significantly eccentric orbit. Its predicted transmission spectroscopy metric (TSM; Sect.~\ref{sec:tsm}) is modest compared to the very highest-priority targets, but still suggests that meaningful constraints may be achievable with a small number of JWST time-series observations, especially when leveraging JWST's broad wavelength coverage to break degeneracies that remain in the optical alone (Sect.~\ref{sec:atmoretrieval}).

We assessed the atmospheric observability of TOI-159\,b, generating equilibrium-chemistry transmission spectra at solar metallicity for three C/O ratios (0.25, 0.5, and 1.0) using \texttt{TauREx\,3}, and propagating these models through a JWST time-series simulator with Gen~TSO \citep{Cubillos2024paspGenTSO}. In our analysis, the most favorable single-visit configuration is NIRSpec/BOTS G395H with the F290LP filter, which delivers high sensitivity across the 3--5~$\mu$m window where H$_2$O and carbon-bearing species (e.g., CO/CO$_2$) can, in principle, provide leverage on the atmospheric molecular inventory and continuum opacity. Figure~\ref{fig:jwst_g395h_three_co} shows the simulated observations for the three C/O scenarios: the predicted uncertainties are comparable to the model-to-model differences, such that the cases largely overlap within the error bars across most of the bandpass. Although this limits our ability to uniquely distinguish C/O within this restricted equilibrium grid using G395H alone, a JWST observation of TOI-159\,b would still deliver valuable constraints giving us information by providing a precise 3--5~$\mu$m baseline that can be combined with shorter-wavelength data to improve retrieval constraints on clouds/hazes and bulk abundances.

Recent JWST observations of HJ atmospheres demonstrate that near-IR time-series spectroscopy can deliver the precision needed to diagnose temperature-dependent chemistry and hemispheric heterogeneity \citep[e.g.,][]{Gapp2025_WASP121b_SiO,Splinter2025_WASP121b_PhaseCurve}. In that context, TOI-159\,b occupies a compelling region of parameter space: a massive, inflated hot Jupiter around a hot, variable star, on an eccentric orbit. The non-zero eccentricity motivates follow-up aimed at the planet's dayside emission and/or phase-dependent behaviour, which can probe radiative timescales and redistribution efficiency.

\end{appendix}
\end{document}